\documentclass[useAMS,usenatbib,usegraphicx]{mn2e}

\newcommand{\hkpc}{h^{-1}{\rm kpc}}
\newcommand{\hmpc}{h^{-1}{\rm Mpc}}

\newcommand{\kms}{\;{\rm km}\,{\rm s}^{-1}}

\newcommand{\vw}{{v_{\rm wind}}}

\newcommand{\msolar}{M_{\odot}}

\newcommand{\gad}{{\sc Gadget-2}}
\newcommand{\CIV}{\hbox{C\,{\sc iv}}}
\newcommand{\OVI}{\hbox{O\,{\sc vi}}}

\title[Mass, Metal, and Energy Feedback in Cosmological
Simulations]{Mass, Metal, and Energy Feedback in Cosmological Simulations}

\author[B. D. Oppenheimer \& R. Dav\'e]{Benjamin D. Oppenheimer$^1$,
Romeel Dav\'e$^1$ \\$^{1}$Astronomy Department, University of Arizona,
Tucson, AZ 85721}

\begin{document}

\pagerange{\pageref{firstpage}--\pageref{lastpage}} \pubyear{2007}

\maketitle

\label{firstpage}

\begin{abstract}

Using \gad~cosmological hydrodynamic simulations including an
observationally-constrained model for galactic outflows, we
investigate how feedback from star formation distributes mass, metals,
and energy on cosmic scales from $z=6\rightarrow 0$.  We include
instantaneous enrichment from Type II supernovae (SNe), as well as
delayed enrichment from Type Ia SNe and stellar (AGB) mass loss, and
we individually track carbon, oxygen, silicon, and iron using the
latest yields.  Following on the successes of the momentum-driven wind
scalings (e.g. Oppenheimer \& Dav\'e 2006), we improve our
implementation by using an on-the-fly galaxy finder to derive wind
properties based on host galaxy masses.  By tracking wind particles in
a suite of simulations, we find: (1) Wind material reaccretes onto a
galaxy (usually the same one it left) on a {\it recycling timescale}
that varies inversely with galaxy mass (e.g. $<1$ Gyr for $L^*$
galaxies at $z=0$).  Hence metals driven into the IGM by galactic
superwinds cannot be assumed to leave their galaxy forever.  Wind
material is typically recycled several times; the median number of
ejections for a given wind particle is 3, so by $z=0$ the total mass
ejected in winds exceeds $0.5\Omega_b$.  (2) The physical distance
winds travel is fairly independent of redshift and galaxy mass ($\sim
60-100$~physical kpc, with a mild increase to lower masses and
redshifts).  For sizable galaxies at later epochs, winds typically do
not escape the galaxy halo, and rain back down in a {\it halo
fountain}.  High-$z$ galaxies enrich a significantly larger comoving
volume of the IGM, with metals migrating back into galaxies to lower
$z$.  (3) The stellar mass of the typical galaxy responsible for every
form of feedback (mass, metal, \& energy) grows by $\sim 30\times$
between $z=6\rightarrow 2$, but only $\sim 2-3\times$ between
$z=2\rightarrow 0$, and is around or below $L^*$ at all epochs.  (4)
The energy imparted into winds scales with $M_{\rm gal}^{1/3}$, and is
roughly near the supernova energy.  Given radiative losses, energy
from another source (such as photons from young stars) may be required
to distribute cosmic metals as observed.  (5) The production of all
four metals tracked is globally dominated by Type II SNe at all
epochs.  However, intracluster gas iron content triples as a result of
non-Type~II sources, and the low-$z$ IGM carbon content is boosted
significantly by AGB feedback.  This is mostly because gas is returned
into the ISM to form one-third more stars by $z=0$, appreciably
enhancing cosmic star formation at $z\la 1$.

\end{abstract}

\begin{keywords}
intergalactic medium, galaxies: abundances, galaxies: evolution, galaxies: high-redshift, cosmology: theory, methods: numerical
\end{keywords}

\section{Introduction}  

Galactic-scale feedback appears to play a central role in the
evolution of galaxies and the intergalactic medium (IGM) over the
history of the Universe.  Mass feedback in the form of galactic
outflows curtails star formation \citep[e.g.][hereafter SH03b]{spr03b}
by removing baryons from sites of star formation, thereby solving the
overcooling problem where too many baryons condense into
stars~\citep[e.g.][]{bal01}.  The energy in these winds carry
metal-enriched galactic interstellar medium (ISM) gas out to large
distances, where the metals are observed in quasar-absorption line
spectra tracing the IGM~\citep[e.g.][]{cow98}.  Galactic outflows
appear to be the only viable method to enrich the IGM to the observed
levels as simulations show that tidal stripping only is not sufficient
\citep[][hereafter OD06]{agu01,opp06}.  Hence understanding galactic
outflows is a key requirement for developing a complete picture of how
baryons in all cosmic phases evolve over time.

Modeling galactic outflows in a cosmological context has now become
possible thanks to increasingly sophisticated algorithms and improving
computational power.  The detailed physics in distributing the
feedback energy from supernovae and massive stars to surrounding gas
still remains far below the resolution limit in such simulations, so
must be incorporated heuristically.  There are two varieties of
approaches of feedback: thermal and kinetic.  \citet{kob04} injects
energy from galactic superwinds and supernovae thermally into a number
of surrounding gas particles in a Smoothed Particle Hydrodynamic (SPH)
simulation, and find that hypernovae with 10$\times$ the typical
supernova energy are needed to enrich the IGM to observed levels while
matching the stellar baryonic content of the local Universe
\citep{kob07}.  \citet{spr03a} (hereafter SH03a) introduced kinetic
feedback in SPH cosmological simulations where individual gas
particles are given a velocity kick and their hydrodynamic forces are
shut off for a period of 30 Myr or until they reach 1/10 the star
formation density threshold.  By converting all the energy from
supernovae into kinetic outflows with constant velocity, SH03b are
able to match the star formation history of the Universe while
enriching the IGM.  \citet{cen06a} introduce a kinetic wind model in
grid-based hydrodynamic simulations, and are able to match the
observed IGM $\OVI$ lines in the local Universe \citep{cen06b}.

In OD06 we took the approach of scaling outflow properties with galaxy
properties, and explored a variety of wind models winds in
\gad~simulations.  We found that the scalings predicted by
momentum-driven galactic superwinds \citep[e.g.][hereafter
MQT05]{mur05} provide the best fit to a variety of quasar absorption
line observations in the IGM, while also reproducing the observed
cosmic star formation history between $z=6\rightarrow1.5$.  In the
momentum-driven wind scenario, radiation pressure from UV photons
generated by massive stars accelerates dust, which is collisionally
coupled to the gas, thereby driving galactic-scale winds.  MQT05
formulated the analytical dependence of momentum-driven winds on the
velocity dispersion of a galaxy, $\sigma$, deriving the relations for
wind velocity, $\vw\propto\sigma$, and the mass loading factor
(i.e. the mass loss rate in winds relative to the star formation
rate), $\eta\propto\sigma^{-1}$.  Observations by \citet{mar05a} and
\citet{rup05} indicate $\vw$ is proportional to circular velocity
(where $v_{\rm circ}\sim\sigma$) over a wide range of galaxies ranging
from dwarf starbursts to ULIRG's.  Mass outflow rates are difficult to
measure owing to the multiphase nature of galactic outflows
\citep{str02,mar05b}, but at least at high-$z$ there are suggestions
that the mass outflow rate is of the order of the star formation rate
in Lyman break galaxies~\citep{erb06}.  A theoretical advantage of
momentum-driven winds is that they do not have the same energy budget
limitations as do supernova (SN) energy-driven winds, where the
maximum is $\sim10^{51}$ ergs per SN, because the UV photon energy
generated over the main sequence lifetime of massive stars is
$\sim100\times$ greater \citep{sch03}.  OD06 found that transforming
all SN energy into kinetic wind energy often is not enough to drive
the required winds, particularly at lower redshifts.  Moreover,
galactic-scale simulations find that in practice only a small fraction
of SN energy is transferred to galactic-scale winds
\citep{mac99,fuj04,fuj08,spi08}.  In short, the momentum-driven wind
scenario seems to match observations of large-scale enrichment, is
broadly consistent with available direct observations of outflows, and
relieves some tension regarding wind energetics.

Still, for the purposes of studying the cosmic metal distribution, the
exact nature of the wind-driving mechanism is not relevant; in our
models, what is relevant is how the wind properties scale with
properties of the host galaxy.  The inverse $\sigma$ dependence of the
mass loading factor appears to be necessary to sufficiently curtail
star formation in high-$z$ galaxies \citep{dav06,fin07a}.  At the same
time, they enrich the IGM to the observed levels through moderate wind
velocities that do not overheat the IGM (OD06).  Continual enrichment
via momentum-driven wind scalings reproduces the relative constancy of
$\Omega(\CIV)$ from $z\approx 6\rightarrow 1.5$ (OD06) and the
approximate amount of metals in the various baryonic phases at all
redshifts~\citep[hereafter DO07]{dav07}.  The observed slope,
amplitude, and scatter of the galaxy mass-metallicity relation at
$z=2$ \citep{erb06} is reproduced by momentum-driven wind
scalings~\citep{fin08}.  While only a modest range of outflow models
were explored in OD06, the success of a single set of outflow scalings
for matching a broad range of observations is compelling.  This
suggest that simulations implementing these scalings approximately
capture the correct cosmic distribution of metals.  Hence such
simulations can be employed to study an important question that has
not previously been explored in cosmological simulations: {\it How do
outflows distribute mass, metals, and energy on cosmic scales?}

In this paper, we explore mass, metallicity, and energy feedback from star
formation-driven galactic outflows over cosmic time.  We use an improved
version of the cosmological hydrodynamic code \gad~\citep{spr05} employing
momentum-driven wind scaling relations, with two major improvements over
what was used in OD06: (1) A more sophisticated metallicity yield model
tracking individual metal species from Type II SNe, Type Ia SNe, and AGB
stars; and (2) An on-the-fly galaxy finder to derive momentum-driven wind
parameters based directly on a galaxy properties.  The OD06 simulations
only tracked one metallicity variable from one source, Type II SNe,
and used the local gravitational potential as a proxy for $\sigma$
in order to determine outflow parameters.  These approximations turn
out to be reasonable down to $z\sim 2$, but at lower redshifts they become
increasingly inaccurate; this was the primary reason why most of
our previous work focused on $z\ga 1.5$ IGM and galaxy properties.
By low-$z$, Type Ia SNe and AGB stars contribute significantly to
cosmic enrichment~\citep{man05, wal98}, and these sources have yields
that depend on metallicity \citep{woo95, lim05}.  Our new simulations
account for these contributions.  Next, using the gravitational potential
wrongly estimates $\sigma$ especially at low-$z$, when galaxies more
often live in groups and clusters and the locally computed potential
does not reflect the galaxy properties alone (as assumed in MQT05).
This tends to overestimate $\vw$ and underestimate $\eta$, resulting
in unphysically large wind speeds and insufficient suppression of star
formation at low redshifts.  Our new simulations identify individual
galaxies during the simulation run, hence allowing wind properties to
be derived in a manner more closely following MQT05.

The paper progresses as follows.  In \S2 we describe in detail our
modifications to \gad, emphasizing the use of observables in
determining our outflow prescription and metallicity modifications.
\S3 examines the energy balance from momentum-driven feedback between
galaxies and the IGM using the new group finder-derived winds.  We
follow the metallicity budget over the history of the Universe in \S4
first by source (\S4.1), and then by location (\S4.2), briefly
comparing our simulations to observables including $\CIV$ in the IGM
and the iron content of the intracluster medium (ICM).  \S5.1 examines
the three forms of feedback (mass, metallicity, and energy) as a
function of galaxy baryonic mass.  We determine the typical galaxy
mass dominating each type of feedback (\S5.2).  We then consider the
cycle of material between galaxies and the IGM, introducing the key
concept of {\it wind recycling} (\S5.3) to differentiate between
outflows that leave a galaxy reaching the IGM and {\it halo fountains}
-- winds that never leave a galactic halo.  We examine wind recycling
as a function of galaxy mass in \S5.4.  \S6 summarizes our results.
We use \citet{and89} for solar abundances throughout; although newer
references exist, these abundances are more easily comparable to
previous works in the literature, and we leave the reader to scale the
abundances to their favored values.

\section{Simulations} \label{sec: sim}

We employ a modified version of the N-body+hydrodynamic code \gad,
which uses a tree-particle-mesh algorithm to compute gravitational
forces on a set of particles, and an entropy-conserving formulation of
SPH \citep{spr02} to simulate pressure forces and shocks in the
baryonic gaseous particles.  This Lagrangian code is fully adaptive in
space and time, allowing simulations with a large dynamic range
necessary to study both high-density regions harboring galaxies and
the low-density IGM.

\gad~also includes physical processes involved in the formation and
evolution of galaxies.  Star-forming gas particles have a subgrid
recipe containing cold clouds embedded in a warm ionized medium to
simulate the processes of evaporation and condensation seen in our own
galaxy \citep{mck77}.  Feedback of mass, energy, and metals from Type
II SNe are returned to a gas particle's warm ISM every timestep it
satisfies the star formation density threshold.  In other words, gas
particles that are {\it eligible} for star formation undergo
instantaneous self-enrichment from Type II SNe.  The instantaneous
recycling approximation of Type II SNe energy to the warm ISM phase
self-regulates star formation resulting in convergence in star
formation rates when looking at higher resolutions (SH03a).

Star formation below 10 $M_{\odot}$ is decoupled from their high mass
counterparts using a Monte Carlo algorithm that spawns star particles.
In \gad~a star particle is an adjustable fraction of the mass of a gas
particle; we set this fraction to 1/2 meaning that each gas particle
can spawn two star particles.  The metallicity of a star particle
remains fixed once formed; however, since Type II SNe enrichment is
continuous while stars are formed stochastically, every star particle
invariably has a non-zero metallicity.  The total star formation rate
is scaled to fit the disk-surface density-star formation rate observed
by \citet{ken98}, where a single free parameter, the star formation
timescale, is set to 2 Gyr for a \citet{sal55} initial mass function
(SH03a).

Even with self-regulation via the subgrid 2-phase ISM, global star
formation rates were found to be too high, meaning another form of
star formation regulation is required.  SH03b added galactic-scale
feedback in the form of kinetic energy added to gas particles at a
proportion relative to their star formation rates.  They set the wind
energy equal to the Type II SNe energy, thereby curtailing the star
formation in order to broadly match the observed global cosmic star
formation history.  SH03b assumed a constant mass loading factor for
the winds, which resulted in a constant wind velocity of 484~km/s
emanating from all galaxies.  OD06 found that scaling the velocities
and mass-loading factors as prescribed by the momentum-driven wind
model did a better job of enriching the high-$z$ IGM as observed,
while better matching the cosmic star formation history.

We have performed a number of modifications to \gad~since OD06.  These
include (1) the tracking of individual metal species, (2)
metallicity-dependent supernova yields, (3) energy and metallicity
feedback from Type Ia SNe, (4) metallicity and mass feedback
from AGB stars at delayed times, (5) a particle group finder to
identify galaxies in situ with \gad~runs so that wind properties can
depend on their parent galaxies, and (6) a slightly modified
implementation of momentum-driven winds.  We describe each in turn in
the upcoming subsections.

\begin{table*}
\caption{Simulation parameters}
\begin{tabular}{lcccccccc}
\hline
Name$^{a}$ &
$L^{b}$ &
$\epsilon^{c}$ &
$m_{\rm SPH}^{d}$ &
$m_{\rm dark}^{d}$ &
$M_{\rm gal,min}^{d,e}$ &
$z_{\rm end}$ &
AGB Feedback? &
Wind Derivation
\\
\hline
\multicolumn {9}{c}{Test Simulations} \\
\hline
l8n128vzw-$\Phi$          & 8  & 1.25 & 4.72   & 29.5 & 151  & 0.0 & Y & $\Phi$    \\
l8n128vzw-$\sigma$        & 8  & 1.25 & 4.72   & 29.5 & 151  & 0.0 & Y & $\sigma$  \\
l8n128vzw-$\sigma$-nagb   & 8  & 1.25 & 4.72   & 29.5 & 151  & 0.0 & N & $\sigma$  \\

l32n128vzw-$\Phi$         & 32 & 5.0  & 302    & 1890 & 9660 & 0.0 & Y & $\Phi$    \\ 
l32n128vzw-$\sigma$       & 32 & 5.0  & 302    & 1890 & 9660 & 0.0 & Y & $\sigma$  \\
l32n128vzw-$\sigma$-nagb  & 32 & 5.0  & 302    & 1890 & 9660 & 0.0 & N & $\sigma$  \\
\hline
\multicolumn {9}{c}{High-Resolution Simulations} \\
\hline
l8n256vzw-$\sigma$        & 8  & 0.625& 0.590  & 3.69 & 18.8  & 3.0 & Y & $\sigma$    \\ 
l16n256vzw-$\sigma$       & 16 & 1.25 & 4.72   & 29.5 & 151   & 1.5 & Y & $\sigma$  \\
l32n256vzw-$\sigma$       & 32 & 2.5  & 37.7   & 236  & 1210  & 0.0 & Y & $\sigma$  \\
l64n256vzw-$\sigma$       & 64 & 5.0  & 302    & 1890 & 9660  & 0.0 & Y & $\sigma$  \\
l64n256vzw-$\sigma$-nagb  & 64 & 5.0  & 302    & 1890 & 9660  & 0.0 & N & $\sigma$  \\
\hline
\end{tabular}
\\
\parbox{15cm}{
$^a$The {\it 'vzw'} suffix refers to the momentum-driven winds with
$L_{\rm f}$ varying between 1.05-2.0, metallicity-dependent $\vw$, and an
extra kick to get out of the potential.\\
$^b$Box length of cubic volume, in comoving $\hmpc$.\\
$^c$Equivalent Plummer gravitational softening length, in comoving
 $\hkpc$.\\
$^d$All masses quoted in units of $10^6M_\odot$.\\
$^e$Minimum resolved galaxy stellar mass.\\
}
\label{table:sims}
\end{table*}

All simulations used here are run with cosmological parameters
consistent with the 3-year WMAP results~\citep{spe07}.  The parameters
are $\Omega_{0} = 0.30$, $\Omega_{\Lambda} = 0.70$, $\Omega_b =
0.048$, $H_0 = 69$ km s$^{-1}$ Mpc$^{-1}$, $\sigma_8 = 0.83$, and $n =
0.95$; we refer to this as the $l$-series.  Note that $\sigma_8$ is
somewhat higher than the WMAP3-favored value of 0.75, owing to
observations suggesting that it may be as high as
0.9~\citep[e.g.][]{roz07,evr08}.  Our general naming convention,
similar to OD06, is l({\it boxsize})n({\it particles/side})vzw-({\it
suffix}) where {\it boxsize} is in $\hmpc$ and the {\it suffix}
specifies how the winds are derived (``$\sigma$" from the on-the-fly
group finder, or ``$\Phi$" from the local gravitational potential) and
whether AGB feedback was {\it not} included (``nagb").

Table \ref{table:sims} lists parameters for our runs presented in this
paper.  We ran a series of test simulations with $2\times128^3$
particles in 8 and 32 $\hmpc$ boxes to explore the effect of turning
off the AGB feedback and using the old prescription of using
potential-derived $\vw$.  The $2\times256^3$ simulations are our
high-resolution simulations and range in gas particle mass from
$0.59-302\times10^6 M_{\odot}$.  The l32n256vzw-$\sigma$ simulation was
by far the most computationally expensive simulation taking in excess
of 50,000 CPU hours on an SGI Altix machine.  The l16n256vzw-$\sigma$
simulation contains the minimum resolution needed to resolve $\CIV$
IGM absorbers (OD06), however it is prohibitively expensive to run
this to $z=0$.  We will use the l8n128vzw simulations at the same
resolution but a smaller box to explore these absorbers; this box
appears to converge with the l16n256vzw simulation at $z=1.5$, despite
containing a smaller volume unable to build larger structures.

\subsection{Metal Yields}

In previous \gad~simulations including SH03b and OD06, metal
enrichment was tracked with only one variable per SPH particle
representing the sum of all metals and was assumed to arise from only
one source, Type II SNe, which enriched instantaneously.  While
this is reasonably accurate when considering oxygen abundances, the
abundances of other species can be significantly affected by alternate
sources of metals.

We have implemented a new yield model that tracks four species (carbon,
oxygen, silicon, and iron) from three sources (Type II SNe, Type Ia SNe,
and AGB stars) all with metallicity-dependent yields.  These sources have
quite different yields that depend significantly on metallicity, and
inject their metals at different times accompanied by a large range in
energy feedback.  A more sophisticated yield model is required to model
metal production from the earliest stars, abundance variations within
and among galaxies, abundance gradients within the IGM, and abundances
in the ICM.

The four species chosen not only make up 78\% of all metals in the
sun \citep{and89}, but are the species most often observed in quasar
absorption line spectra probing the IGM, X-ray spectra of the ICM,
and the ISM of galaxies used to determine the galaxy mass-metallicity
relationship.  Furthermore, because these metals are among the most
abundant, they are also often the most dynamically important when
considering metal production in stars, the multi-phase ISM, and metal-line
cooling of the IGM.  We have not implemented metal-line cooling per
individual species, but this may be straightforwardly incorporated in
the future.

\subsubsection{Type II Supernovae}

Type II SNe enrichment follows that presented in SH03a, namely their
equation~40 where gas particles are self-enriched instantaneously via
\begin{equation}\label{eqn:TypeIISN}
\Delta Z_{\rm species} = (1 - f_{\rm SN}) y_{\rm species}(Z) x \frac{\delta t}{t_*}
\end{equation}
where $f_{\rm SN}$ is the fraction of the stellar initial mass
function (IMF) that goes supernova, $x$ is the fraction of an SPH gas
particle in the cold ISM phase, and $t_{*}$ is the star formation
timescale.  Our modification is that we follow the yield of each
species individually using metallicity-dependent yields, $y_{\rm
species}(Z)$, from the nucleosynthetic calculations by \citet{lim05}
instead of assuming $y=Z_{\odot}=0.02$ as SH03b and OD06 did.  Their
grid of models include SNe ranging from 13 to 35 $M_{\odot}$ and
metallicities ranging from $Z=0-0.02$.  Using the total metallicity of
a gas particle (i.e. the sum of the four species divided by 0.78 to
account for other species), we employ a lookup table indexed by
metallicity to obtain the $y_{\rm species}(Z)$. The \citet{lim05}
yields are the most complete set of metallicity-dependent yields since
\citet{woo95}.  Both papers find similar yields for carbon and oxygen,
the two species most important for IGM observations.

We use the \citet{cha03} IMF, although it is quite possible to modify
the IMF in the future so as to have a top-heavy IMF under different
conditions~\citep[e.g.][]{dav08a}.  We assume all stars between 10-100
$M_{\odot}$ go supernova, comprising $f_{\rm SN} = 0.198$ (i.e. 19.8\%
of the total stellar mass in the IMF).  We use the yields of
\citet{lim05} comprising $13-35 M_{\odot}$ over this larger mass
range, thus assuming the similar yields from stars between $10-13
M_{\odot}$ and $35-100 M_{\odot}$.  Other supernova yield models that
include more massive stars \citep{por98,hir05} show higher carbon and
oxygen yields from stars over 40 $M_{\odot}$ at solar metallicity, but
do not cover the range of metallicities of \citet{lim05}.  The
fraction of stars going supernova, $f_{\rm SN}$, is nearly twice as
high since SH03a assumes a Salpeter IMF ($f_{\rm SN}=0.1$), because
there is a turnover at masses less than 1 $M_{\odot}$.  The remaining
80.2\% form star particles from which AGB feedback arises at later
times.

\subsubsection{Type Ia Supernovae} \label{sec: typeia}


Type Ia SNe are believed to arise from mass accretion from a companion
star that increases the mass beyond the Chandrasekhar limit, causing
an explosion.  Recently, the Type Ia SNe rate was measured by
\citet{man05}, and the resulting data was parameterized by
\citet{sca05} with a two-component model, where one component is
proportional to the stellar mass (``slow" component), and the other to
the star formation rate (``rapid" component):
\begin{equation}\label{eqn:typeIa}
SNR_{\rm Ia} = A M_\star + B \dot{M_\star} \;\;\; M_\odot/{\rm yr}
\end{equation}
\citet{sca05} determined that the best fit to the \citet{man05} data
was provided by $A=4.4\times 10^{-14}$~yr$^{-1}$ and $B=2.6\times
10^{-3}$, with a time delay of 0.7~Gyr in the slow component for the
onset of Type~Ia SNe production.

To implement this in \gad, we calculate the number of Type Ia SNe
formed at each timestep for every gas particle (from the first part of
eqn.~\ref{eqn:typeIa}) and every star particle where the star was
formed more than 0.7~Gyr ago (from the second term in
eqn.~\ref{eqn:typeIa}).  Each Type~Ia SN is assumed to add
$10^{51}$~ergs of energy, which is added directly to the gas particle
or, in the case of star particles, added to the nearest gas particle.
Each Type~Ia is also assumed to produce $0.05 M_{\odot}$ of carbon,
$0.143 M_{\odot}$ of oxygen, $0.150 M_{\odot}$ of silicon, and $0.613
M_{\odot}$ of iron with these four species making up 78\% of all
metals made by Type Ia SNe \citep{thi86, tsu95}.

\subsubsection{AGB Stars} \label{sec: AGBfeedback}

Feedback from AGB stars comprise at least half of the total mass
returned to the ISM \citep{wal98}.  AGB stars copiously produce
carbon, referred to as carbon stars, and other isotopes of carbon,
nitrogen, and oxygen \citep{ren81} on a delayed timescale compared to
the relatively instantaneous enrichment of Type II SNe.  Heavier
elements such as silicon and iron that remain unprocessed by low-mass
stars can now be returned back into the ISM instead of being trapped
in stars.  The mass and metallicity feedback from AGB stars is
considerable and in some regions even dominant over supernovae.
However, relative to supernovae, the energy feedback can be considered
negligible, since most of the mass leaves AGB stars at far less than
100~$\kms$.  


First we consider the mass feedback as a function of time for a star
particle.  We use the \citet{bru03} stellar synthesis models using the
Padova 1994 \citep{bre93} libraries of stellar evolution tracks to
determine the mass loss rate from non-supernova stars as a function of
age given a Chabrier IMF.  Mass loss rates are calculated at an age
resolution of 0.02 dex for six different metallicities ($Z$=0.0001,
0.0004, 0.004, 0.008, 0.02, and 0.05) covering stellar ages from
$log(t)=7.18-10.14$ (i.e. the age of the death of the most massive AGB
stars to the age of the Universe).  We interpolate in age and
metallicity.

To implement this in \gad~we use each star particle's age and
metallicity to determine the mass loss rate from a lookup table
generated from the \citet{bru03} population synthesis models.  This is
performed for every star particle each timestep with the total amount
of mass lost being the product of the timestep and the mass loss rate.
This mass is then transferred to the 3 nearest gas particle neighbors
using a neighbor search.

To illustrate the importance of mass feedback from low-mass stars
consider a stellar population at $Z=0.02$, which by $log(t)=7.34$ yrs
has returned about 17.5\% of its mass via supernova feedback.  By 100
Myr another 10.3\% of the mass is returned to the ISM, and another
12.0\% is returned by 1 Gyr.  More mass is returned to the ISM in the
first Gyr from AGB stars than SNe.  Another 9.7\% is returned between
1 and 10 Gyr.  Overall, slightly greater than 50\% of the mass of a
Chabrier IMF is returned to the ISM with more than 30\% coming from
low and intermediate mass stars.

To model the return of metals into the ISM, we use the stellar yield
models of \citet{her04} ($Z=0.0001$), \citet{mar01} ($Z=$0.008), and
\citet{gav05} ($Z=$0.0126, 0.0200, 0.0317) for a variety of stellar
masses.  We take the age at which a given stellar mass ends its life,
from which we generate an interpolated lookup table of yields for a
given species analogous to the mass loss lookup table.  The
incremental parcel of mass lost during each time step for each star
particle is given a yield corresponding to the mass of a star dying at
the star particle's age.  This assumes that only one specific stellar
mass is contributing to the entire mass loss from a stellar population
at a given age, which is not a bad assumption considering that
intermediate mass stars lose most of their mass during the short-lived
AGB stage.

\begin{figure}
\includegraphics[scale=0.80]{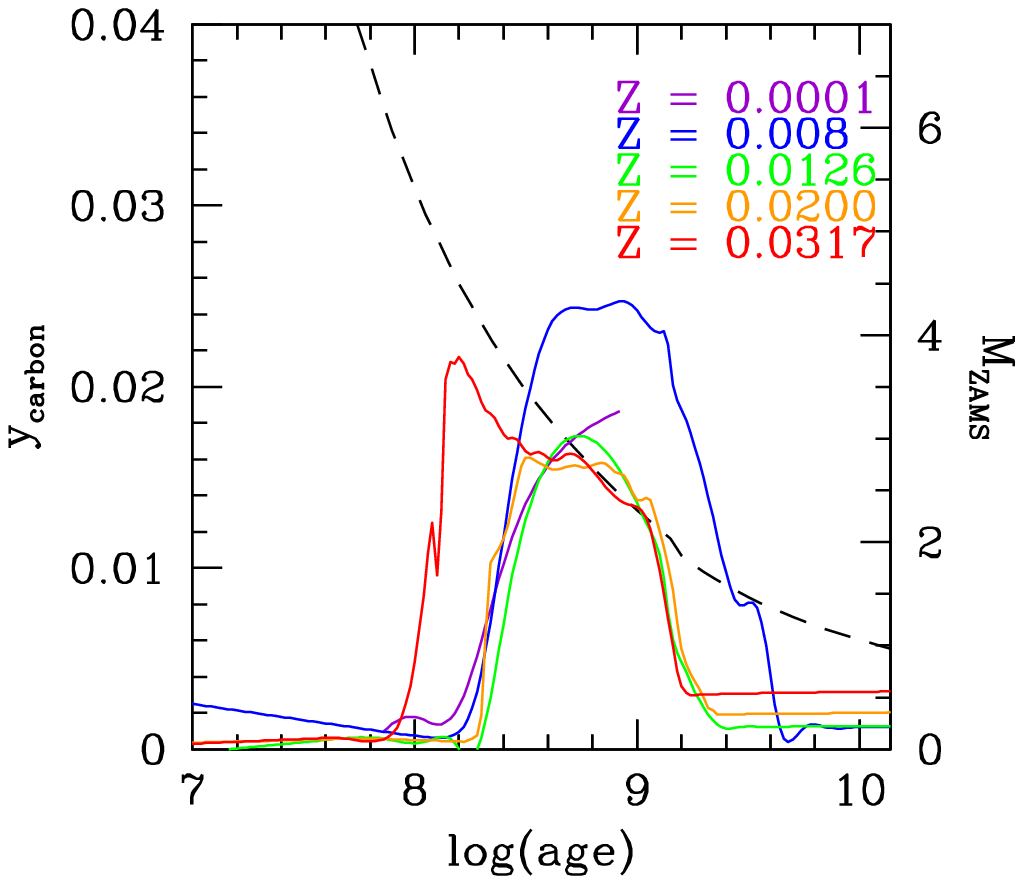}
\caption[]{Carbon yields of gas lost from AGB stars for several
metallicities calculated by \citet{her04}, \citet{mar01} and
\citet{gav05} converted from a function of stellar mass to a function
of age by using a star's lifetime (dashed line).  In \gad, we take a
star particle's age and assign the yield from these curves to the
fractional amount of mass lost from the star particle and added to the
3 nearest SPH neighbors.  The carbon yields jump between $\sim200$ Myr
and 1 Gyr when AGB stars in the range of 2-4 $M_{\odot}$ are dying as
carbon stars; AGB yields are highly dependent on age as well as
metallicity.  The $Z=0.008$ yields show the strongest yields,
especially toward lower mass, while by $Z=0.0317$ the strongest yields
come from $5M_{\odot}$ stars.}
\label{fig:cyield}

\end{figure}

We calculate AGB yield lookup tables for carbon and oxygen only.
Silicon and iron remain almost completely unprocessed in low and
intermediate mass stars, so therefore we simply take their yields to
be the original abundances of the star particle when it formed.  To
illustrate the yields, Figure~\ref{fig:cyield} plots the carbon yield
as a function of age for a variety of metallicities.  The dashed line
is the relation between the Zero-age Main Sequence mass and the age of
death.  Carbon stars enrich copiously between $\sim$200 Myr and 1 Gyr
corresponding to stars of masses 2-4 $M_{\odot}$ going through the AGB
stage.  The reason for this is that third dredge-up becomes efficient
above 2 $M_{\odot}$ transporting the products of double shell burning
into the envelope (i.e. C$^{12}$) until hot-bottom burning becomes
efficient above 4-5 $M_{\odot}$ transforming carbon into nitrogen.
\citet{lia02} also added delayed AGB feedback into SPH simulations,
although they do not extract yields directly from models; instead they
use a time-dependent yield function without metallicity dependence for
carbon.

It is worth pointing out that our simulation do not track the metals
in stellar remnants.  The metal products arising from nucleosynthesis
in AGB stars for the most part remain in the white dwarf remnant once
it has blown its envelope off, and we do not track the creation of
these metals in the star particles.  The same is true for neutron
stars and stellar black holes.  Fortunately these metals remain locked
in their stellar remnants on timescales far beyond the age of the
Universe, hence we can ignore them as a component of observable
metals.  When we talk about the global metal budgets in \S\ref{sec:
Zbudget}, we do not include metals trapped in stellar remnants.

\subsection{Group Finder} \label{sec: groupfinder}

A group finder added to \gad~allows us to add new dynamics based on the
properties of a gas or star particle's parent galaxy.  This is especially
important for momentum-driven winds, which MQT05 argued depend on the
properties of a galaxy as a whole (specifically a galaxy's $\sigma$).
We will show in \S\ref{sec: windmodel} that winds derived from the group
finder more accurately determine wind speeds for the momentum-driven wind
model versus using the potential well depth of a particle as a proxy of
galaxy mass as was done in OD06 and DO07.

Our group finder is based on the friends-of-friends (FOF) group finder
kindly provided to us by V. Springel, which we modified and parallelized
to run in situ with \gad.  Gas and star particles within a specified
search radius are grouped together and cataloged if they are above
a certain mass limit, which we set to be 16 gas particle masses.
The search radius is set to 0.04 of the mean interparticle spacing
corresponding to an overdensity of $(1/0.04)^3 = 15625$.  We chose
this value after numerous comparisons with the more sophisticated (but
far more computationally expensive) Spline Kernel Interpolative DENMAX
(SKID)\footnote{http://www-hpcc.astro.washington.edu/tools/skid.html}
group finder performed on snapshot outputs.

The FOF group finder is run every third domain decomposition, which
corresponds to an interval usually between 2 and 10 Myr depending on
the dynamical time-stepping for the given simulation (i.e. smaller
timesteps for higher resolution).  This timestep was chosen to be
smaller than the dynamical timescales of most if not all galaxies
resolved in a particular simulation.  The galaxy properties output
include gas and stellar mass and metallicities, star formation rate,
and fraction of gas undergoing star formation.  The additional CPU
cost ranges from 7-10\% for a 64-processor run to less than 5\% for a
16-processor run.  Once the FOF group finder is run, each particle in
a galaxy is assigned an ID so that it can be easily linked to all the
properties of its parent galaxy.

\begin{figure*}
\includegraphics[scale=0.80]{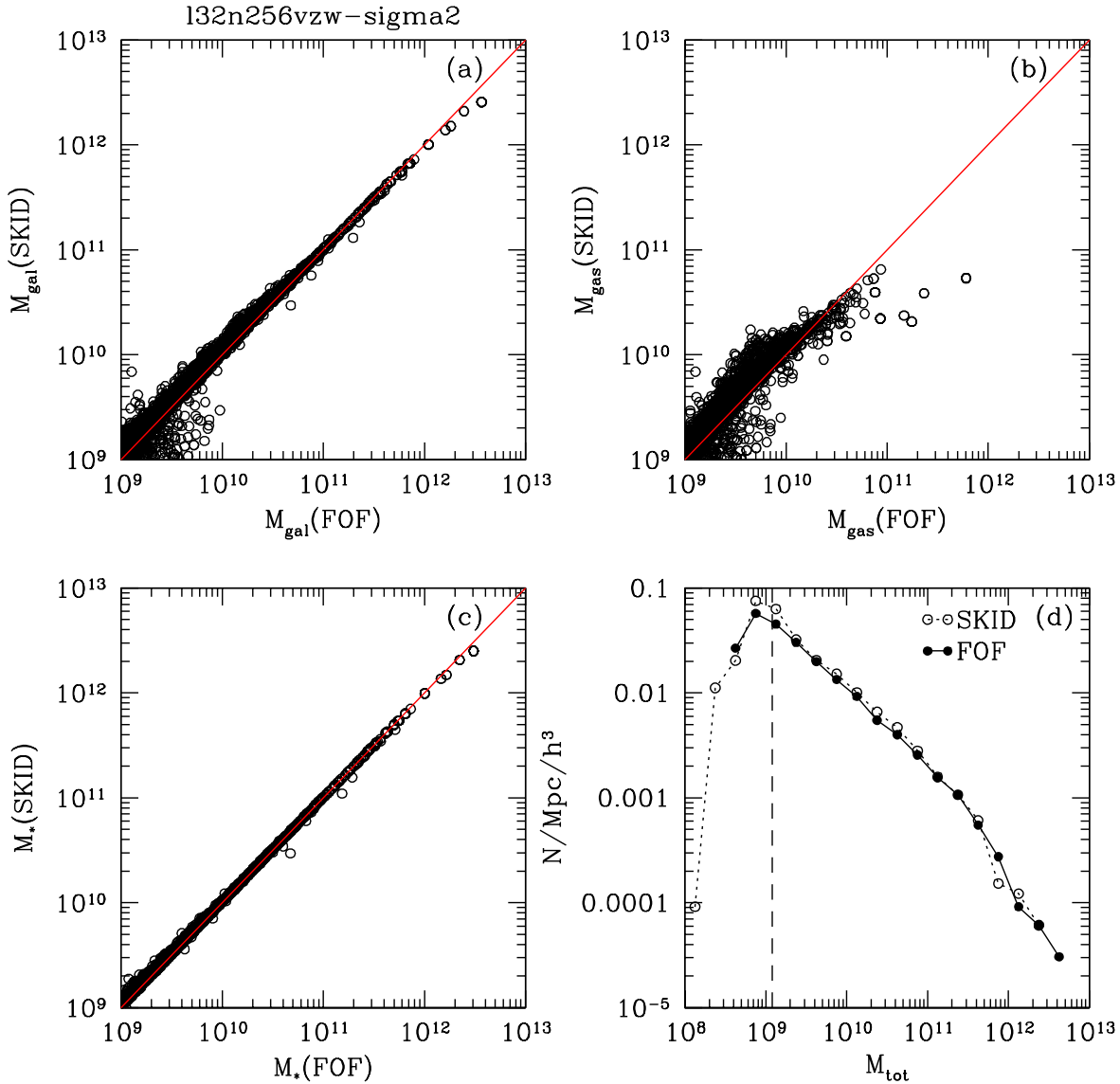}
\caption[]{The comparison of two group finders run on a 32 $\hmpc$
$2\times256^3$ simulation at $z=0$.  FOF is the much faster algorithm
run in situ with \gad~at frequent intervals to determine galaxy
properties on the fly, while SKID is a post-processing DENMAX
algorithm run on simulation snapshots.  Panels (a), (b), and (c)
compare the masses (total, gas, and stellar mass respectively) of
groups found within 5 kpc of each other by the two group finders.
Stellar masses agree very well while the gas masses (a smaller
component by $z=0$) show more scatter.  The total mass functions
(Panel (d)) of both group finders lie nearly on top of each other
above the resolution limit (dashed line), although FOF tends to group
some satellites found by SKID with the largest few galaxies.  The
$M_{\rm gal}$ functions do not resemble a Schechter function, although
when one plots $M_{*}$ there is a more pronounced turnover below
$M_*=10^{10} M_{\odot}$.  }
\label{fig:skidfofcomp}
\end{figure*}

A comparison plot of the two group finders in
Figure~\ref{fig:skidfofcomp} for the l32n256vzw-$\sigma$ simulation at
$z=0$ shows very good but not perfect agreement for total baryonic
galaxy mass ($M_{\rm gal}$, Panel (a)).  Galaxies are matched up by
requiring a $<5\hkpc$ difference between SKID and FOF positions.
Stellar masses ($M_*$, lower left panel) are nearly identical, but the
associated gas mass ($M_{\rm gas}$, upper right) shows more scatter.
Note that there is no explicit density or temperature threshold for
including gas in the FOF case, but in the SKID case only gas with
overdensities $>1000$ compared to the cosmic mean are included; this
may contribute to some of the scatter.  FOF group finders have the
tendency to group too many things together in dense environments, and
this is most noticeable in the associated gas masses.  There are
significantly better agreements in gas masses at higher redshift,
where dense group/cluster environments are less common and gas
fractions are greater.

The mass functions of both group finders when all galaxies
are included show very good agreement (lower right panel
of Figure~\ref{fig:skidfofcomp}).  SKID and FOF each find 6
galaxies with $M_{\rm gal}>10^{12} M_{\odot}$, 112 and 113 galaxies
with $10^{12}>M_{\rm gal}>10^{11} M_{\odot}$, 793 and 719 galaxies
with $10^{11}>M_{\rm gal}>10^{10} M_{\odot}$, and 4296 and 3596
$10^{10}>M_{\rm gal}>10^{9} M_{\odot}$ respectively.  $1.2\times10^{9}
M_{\odot}$ is the galaxy mass resolution limit defined as the mass of 64
SPH particles~\citep{fin06}.  Owing for the tendency of FOF to over-group
satellite galaxies with central galaxies in dense environments, there
is a deficiency of small satellite galaxies in the FOF case.  The total
amount of mass in all resolved galaxies is within 0.2\% between the two
group finders, however there is 24\% more mass grouped to the 6 largest
galaxies in the FOF finder.

As a side note, our group finder has the flexibility to enable
modeling of merger-driven or mass-threshold processes such as AGN
feedback.  Although our simulations implicitly include both hot and
cold-mode accretion \citep{ker05}, \citet{dek06} suggests AGN feedback
affects only hot mode accretion, and hence there exists a threshold
halo mass above which AGN feedback is effective.  Our group finder can
identify galaxies where AGN feedback may be necessary to curtail star
formation and drive AGN winds.  Conversely, if AGN feedback is driven
by the onset of a merger~\citep{dim05}, our group finder can identify
mergers and add merger-driven AGN feedback that curtails star
formation.  We leave such implementations of AGN feedback for future
work, though we note that the heating of gas may transport a
non-trivial amount of energy and metals into the IGM.

Frequent outputs of the group finder allow us to track the formation
history of galaxies to see how galaxies build their mass (e.g. through
accretion or mergers).  We can trace the star formation rate during
individual mergers to see if our simulations are producing bursts of
star formation.  Outputting group finder statistics during a run allows
us to immediately look at integrated properties such as the galaxy
mass function, the mass-metallicity relationship, and the specific
star formation rate as a function of mass just to name a few.  This is
valuable as these relations may exhibit changes on timescales shorter
than the simulation snapshot output frequency.

\subsection{Wind Model Modifications} \label{sec: windmodel}

We use the same kinetic wind implementation of SH03a, whereby
particles enter the wind at a probability $\eta$, the mass loading
factor relative to the star formation rate.  Wind particles are given
a velocity kick, $\vw$, in a direction given by ${\bf v}\times{\bf a}$
(ie. perpendicular to the disk in a disk galaxy).  These particles are
not allowed to interact hydrodynamically until either they reach a SPH
density less than 10\% of the star formation density threshold or the
time it takes to travel 30 kpc at $\vw$; the first case overwhelms the
instances of the second case in our simulations.  We admit that this
phenomenological wind implementation insufficiently accounts for the
true physics of driving superwinds as well as the multi-phase aspect
of winds \citep{str02,mar05b}; see \citet{dal08} for an in-depth
discussion of some of the insufficiencies of such winds in
simulations.  However we want to stress that while we cannot hope to
model the complexities of the outflows, the focus of this paper
primarily depends on the much longer period of subsequent evolution.
We use the momentum-driven wind model with variable $\vw$ and $\eta$,
due to its successes in previous publications mentioned earlier.


In the momentum-driven wind model analytically derived by MQT05, $\vw$
scales linearly with the galaxy velocity dispersion, $\sigma$, and
$\eta$ scales inversely linearly with $\sigma$.  We again use the
following relations:
 \begin{eqnarray}
  \vw &=& 3\sigma \sqrt{f_L-1}, \label{eqn: windspeed} \\
  \eta &=& {\sigma_0\over \sigma} \label{eqn: massload},
 \end{eqnarray} 
where $f_L$ is the luminosity factor in units of the galactic
Eddington luminosity (i.e. the critical luminosity necessary to expel
gas from the galaxy), and $\sigma_0$ is the normalization of the mass
loading factor.  The outflow models used in this paper are all of the
{\it 'vzw'} variety described in \S2.4 of OD06, in which we randomly
select luminosity factors between $f_{\rm L}=1.05-2.00$, include a
metallicity dependence for $f_{\rm L}$ owing to more UV photons output by
lower-metallicity stellar populations
\begin{equation}
  f_L = f_{\rm{L},\odot} \times 10^{-0.0029*(\log{Z}+9)^{2.5} + 0.417694},
  \label{eqn: zfact}
\end{equation}
and add an extra kick to get out of the potential of the galaxy in
order to simulate continuous pumping of gas until it is well into the
galactic halo (as argued by MQT05).

One difference is that we use $\sigma_0=150 \kms$ instead of 300
$\kms$ in the equation for $\eta=\sigma_{0}/\sigma\times$SFR.  Since
our assumed WMAP-3 cosmology produces less early structure that the
WMAP-1 cosmology used in OD06, it requires less suppression of early
star formation and hence lower mass loading factors~(DO07).  We find
that $\sigma_0=150 \kms$ with the WMAP-3 cosmology generally
reproduces the successes of $\sigma_0=300 \kms$ with the WMAP-1
cosmology.

The main modification we make to the outflow implementation is how
$\sigma$ is derived.  Putting the wind parameters in terms of $\sigma$
is the most natural, because the fundamental quantity MQT05 use to derive
momentum-driven winds is $\sqrt{GM\over r}$, which equals $\sqrt{2}
\sigma$ for an isothermal sphere.  In previous runs without a group
finder, we derived $\sigma$ using the virial theorem where $\sigma =
\sqrt{-{1\over 2}\Phi}$, with $\Phi$ being the gravitational potential
at the initial launch position of the wind particle.  We will call this
old form of the wind model $\Phi$-derived winds.  While this derivation
of $\sigma$ should adequately work for an isothermal sphere, the $\Phi$
calculated by \gad~is the entire potential: the galaxy potential on
top of any group/cluster potential.  As galaxies live in groups and
clusters more often low redshift, the wind speeds from $\Phi$-derived
winds become overestimated.  To counteract this trend, we artificially
implemented a limit of $\sigma = 266$ $\kms$, which corresponds to a
$M_{\rm gal}\sim 2.7\times10^{12}$ $\msolar$, the mass of a giant elliptical
galaxy at $z=0$.  In the deep potential of a cluster, all galaxies no
matter what size would drive winds at the speed of this upper limit.
The overestimate of $\vw$ prevented us from trusting our wind model at
lower redshifts; therefore we usually stopped our simulations in OD06
at $z=1.5$.

We now introduce $\sigma$-derived winds, where $\sigma$ is calculated
from $M_{\rm gal}$ as determined by the group finder.  We use the same
relation as MQT05 (their equation~6) from \citet{mo98} assuming the
virial theorem for an isothermal sphere:
\begin{equation} \label{eqn: sigma}
  \sigma = 200 \left(\frac{M_{\rm gal}}{5\times10^{12}} \frac{\Omega_m}{\Omega_b} h \frac{H(z)}{H_{0}}\right)^{1/3} \kms.
\end{equation}
Since our group finder links only baryonic mass, we multiply $M_{\rm gal}$
by ${\Omega_m}/{\Omega_b}$ to convert to a dynamical mass.  The wind
properties are hence estimated from the galaxy alone.  We do not 
calculate $\sigma$ from the velocity dispersion of star and/or
gas particles in the galaxy, since our tests show the resolution is
insufficient to derive a meaningful $\sigma$.

The $H(z)/H_0$ factor in equation~\ref{eqn: sigma} increases $\sigma$
for a given mass as $H(z)$ increases at higher redshift.  For example,
$H(z=6) = 10.2\times H_0$ resulting in winds being $2.17\times$ as fast
being driven from the same mass galaxy at $z=6$ versus $z=0$.  Physically,
galaxies that form out of density perturbations at higher $z$ have
overcome faster Hubble expansion, and therefore their $\sigma$ is higher.
The same mass galaxy formed at higher redshift with a higher $\sigma$
means the higher redshift galaxy must be more compact.  Such a scenario
is supported by the observations by \citet{tru06} and \citet{tru07}
showing a trend of galaxies becoming smaller from $z=0\rightarrow2$,
in general agreement with \citet{mo98}.  A more compact galaxy drives a
faster wind in the momentum-driven wind model, because $\vw$ is a function
of $M_{\rm gal}/r$, not just $M_{\rm gal}$; there is more UV photon flux impinging
each dust particle.  We will show that increasing $\vw$ emanating from
the same mass galaxy toward higher $z$ has important consequences in
enriching the IGM at high-$z$ while not overheating it at low-$z$.

Another modification to the wind model we add are new $\vw$ speed
limits.  The first depends on the star formation timescale,
$\tau_{\rm SFR}$.  The momentum-driven wind equation derived by MQT05 and
used in OD06 and DO07 assumes a starburst occurs on the order of a
dynamical timescale, $\tau_D$, which is often the case in a
merger-driven starburst.  However, MQT05 also derive a maximum
$\sigma$, $\sigma_{\rm max}$, above which a starburst cannot achieve the
luminosity needed to expel the gas in an optically thick case.
Although it is not clear how $\sigma_{\rm max}$ varies with the
$\tau_{\rm SFR}$, we modify their equation 23 to have an inverse linear
dependence on the star formation timescale
\begin{equation}
  \sigma_{\rm max} = 4000 \times \frac{\tau_{\rm D}}{\tau_{\rm SFR}} \kms
\end{equation}
and assume a $\tau_D$ of 50 Myr.  The end result is a reduction of
5-10\% in the average $\vw$ out of $M_{\rm gal}\sim10^{12}$ at $z<1$.

A second speed limit we impose allows no more than $2\times$ the total
SN energy to be deposited into the wind.  This does not violate the
energetics as the energy for these winds is coming from momentum
deposition over the entire lifetime of a star, which is of the order
$100\times$ the SN energy \citep{sch03}.  This limit was instituted to
disallow extreme values of $\vw$ (i.e. $>1500 \kms$) emanating from
the most massive haloes, reducing $\vw$ at most 30-50\% in the most
massive galaxies at $z<1$.

\begin{figure}
\includegraphics[scale=0.80]{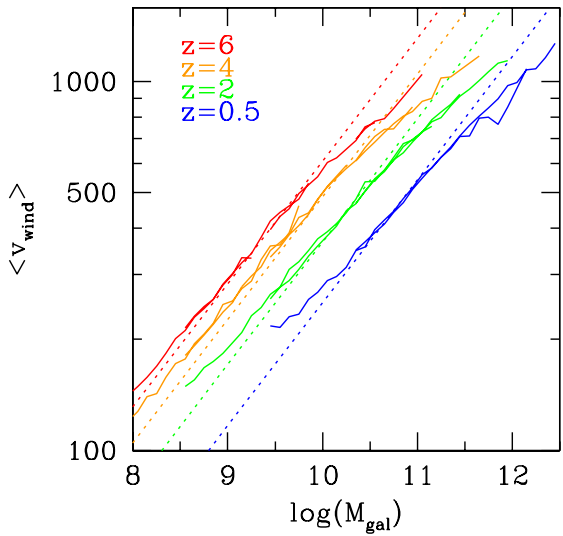}
\caption[]{Average launch wind speed as a function of $M_{\rm gal}$
(derived using SKID) versus dotted lines for the predicted
$\langle\vw\rangle$ assuming no speed limits.  The wind speeds, which
are derived using the FOF finder during a run, agree very well with
the masses determined by SKID in the post-processing stage.  The
decline in $\langle\vw\rangle$ with redshift for a given $M_{\rm gal}$ is
a result of the $H(z)$ dependence in equation \ref{eqn: sigma}.  This
trend loosely matches observations of faster winds for a given mass
galaxy at high redshift.  }
\label{fig:vwind}
\end{figure}

In Figure~\ref{fig:vwind}, we plot the average wind speed,
$\langle\vw\rangle$, as a function of galaxy mass at four redshifts
for a variety of box sizes ranging from 8 to 64 $\hmpc$ along.  Dotted
lines show the predicted $\langle\vw\rangle$ for solar metallicity,
assuming no speed limits.  As with every plot in this paper, $M_{\rm
gal}$ is the SKID-derived baryonic mass, not the FOF-derived mass from
which $\vw$ is calculated.  The simulations reproduce the predicted
trend $\vw\propto M_{\rm gal}^{1/3}$.  Divergences at low mass result
from faster winds being driven by low-mass, low-metallicity galaxies
as well as some satellites being grouped with a central galaxy in
groups/clusters by FOF.  The latter effect appears to be sub-dominant
though as evidenced by the overlap of the relation among simulations
at different redshifts.  The deviations at the high-mass end at all
redshifts are dominated by the second wind speed limit discussed
above.

The reduction in $\vw$ for a given mass galaxy due to the $H(z)$
dependence in equation \ref{eqn: sigma} is the reverse of the trend in
the $\Phi$-derived models, and are in better agreement with low-$z$
observations.  In the new $\sigma$-derived wind model, a $10^{11}
M_{\odot}$ galaxy (such as the Milky Way) launches an average wind
particle at 790 $\kms$ wind at $z=3$ and 450 $\kms$ at $z=0$ while the
corresponding values for the $\Phi$-derived winds without any speed
limits are 1040 $\kms$ and 1220 $\kms$ respectively.  The latter are
far above the values observed in the local Universe.  \citet{mar05a}
found the relation $v_{\rm term}\propto 2.1\times v_{\rm circ}$ fit her
observations best, where $v_{\rm term}$ is the terminal velocity of the
wind.  For a $10^{11} M_{\odot}$ galaxy, this corresponds to
$v_{\rm circ}\sim125$ $\kms$ leading to $v_{\rm term}\sim 265$ $\kms$.  This
is much nearer our $\sigma$-derived value once the extra velocity
boost to leave the potential of the galaxy is subtracted.
High-velocity clouds (HVC's) may be material that is blown into the
halo from the Milky Way~\citep{wak97}, and these have velocities that
are $\ll 1000\;\kms$, in better agreement with velocities expected in
the $\sigma$-derived model.  Of course the Milky Way is not a
star-bursting galaxy driving a powerful wind; however, it is still
possible that it is kicking up a significant amount of gas into the
halo.  Indeed, as we will discuss later (\S\ref{sec: galrecycle}), the
outflows in our simulations don't always reach the IGM, but
particularly at low-$z$ may be more aptly described as ``halo
fountains", where the outflows only propagate out to distances
comparable to the galactic halo.

It remains very difficult to determine from observations whether wind
speeds increase for a given mass galaxy at high-$z$ as the
$\sigma$-derived winds predict.  Observing asymmetric Lyman-$\alpha$
profiles,\citet{wil05} find what they interpret as 290 $\kms$ outflows
around a LBG with $10^{11} M_{\odot}$ baryons at $z=3.09$.  More
recently \citet{swi07} observe outflows up to 500 $\kms$ at $z=4.88$
around a lensed galaxy with a dynamical mass as low as $10^{10}
M_{\odot}$.  These observed outflows appear to be beyond the virial
radius in both cases and correspond to our velocities once we have
subtracted the extra velocity boost we add to get out of the potential
well.  Our $v_{\rm term}$ for the \citet{wil05} $z=3$ galaxy would be
closer to $500 \kms$, somewhat above their observed values.  We expect
$\vw\sim300\kms$ ($v_{\rm term}\sim200$ $\kms$) for the \citet{swi07}
object, but their mass is only a lower limit.  If this object is a
$10^{10} M_{\odot}$ baryonic-mass galaxy, we would derive
$v_{\rm term}\sim400\kms$.  Overall, the $\sigma$-derived winds predict
velocities that are at least in the ballpark of observed values.
Additionally, surveys of LBG's at$z\sim3$ \citep{pet01, sha03} find
outflows of several hundred $\kms$ to be ubiquitous, often driving an
amount of mass comparable to the star forming mas (i.e. $\eta\sim1$)

\section{The Universal Energy Balance} \label{sec: globalstats}

Armed with these new simulations, we can now investigate how outflows move
mass, metals, and energy around the Universe.  In this section we focus
on the energetics of outflows, and its impact on cosmic star formation
and temperature.  We will compare the new $\sigma$-derived wind model's
behavior versus the old $\Phi$-derived wind model, and study the impact
of AGB feedback. Specifically, $\sigma$-derived winds inject much less
energy at late times making a cooler and less-enriched IGM while leading
to more star formation.  The inclusion of AGB feedback does not really
affect the global energy balance, but does increase the number of stars
formed and more significantly affects the amount and location of metals,
as we discuss in \S\ref{sec: Zbudget}.

\begin{figure*}
\includegraphics[scale=0.8]{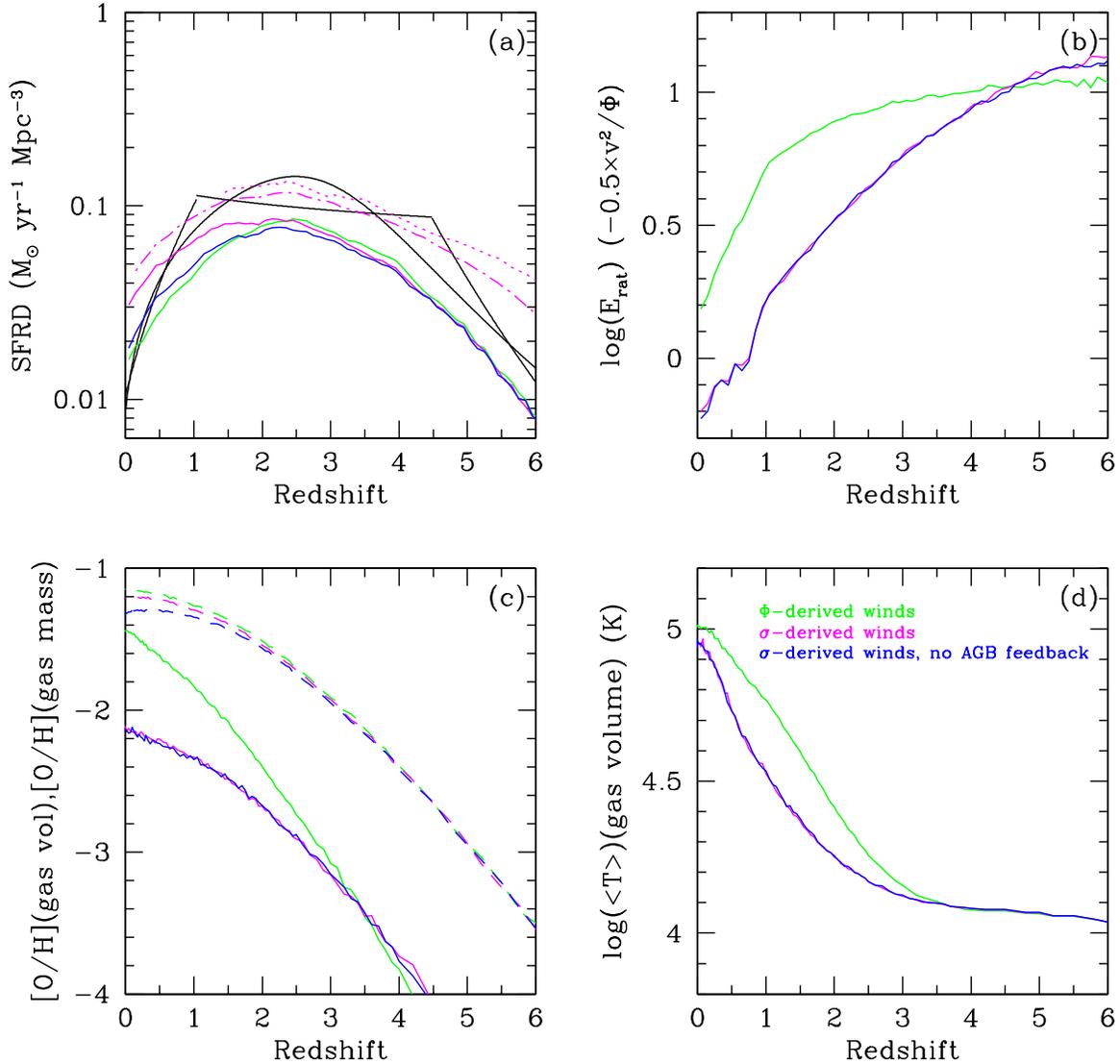}
\caption[]{Three 32 $\hmpc$ $2\times128^3$ simulations: $\Phi$-derived
winds including AGB feedback (green), $\sigma$-derived winds with AGB
feedback (magenta), and $\sigma$-derived winds without AGB feedback
(blue) demonstrate how global quantities evolve as a result of
energetic interactions between star formation-driven feedback from
galaxies and the state of the IGM.  Panel (a) shows the star formation
density histories compared to the \citet{hop06} compilation averages
(black lines for segmented and non-segmented fit, scaled to a Chabrier
IMF).  Panel (b) shows the average virial ratio with which wind
particles are launched.  Solid/dashed lines in Panel (c) trace the
volume/mass-weighted average gas oxygen content, and Panel (d) shows
the volume-weighted temperature.  The new $\sigma$-derived winds do
not inject as much energy or metals into the IGM while increasing the
star formation.  AGB feedback has a relatively minor effect in the
energetic balance between galaxies and the IGM.  The lower resolution
l32n128 simulation misses about half of all star formation, therefore
we also show the l32n256 (dash-dotted line) and the l16n256 (dotted
line) simulations in Panel (a), which both agree better with
observations. }
\label{fig:global}
\end{figure*}

The history of the cosmic star formation rate
density~\citep[SFRD;][]{mad96,lil96} is a key observable that has
received much attention in recent years~\citep[e.g.][]{hop06,far07}.
The SFRD plot in the upper left panel of Figure~\ref{fig:global} shows
how our models compare to the \citet{hop06} compilation (black lines,
two different fits).  It should be noted that we are resolving
galaxies down to $\approx 10^{10} M_{\odot}$ in the models shown with
solid lines, and hence the star formation density at $z\ga 3$ is
increasingly underestimated (see e.g. SH03b).  Hence this should be
considered as an illustrative comparison, whose main point is to
highlight the differences between various models.  A more complete SF
history at early times is shown by the dashed and dotted magenta lines
from our higher resolution $2\times256^3$ particle simulations with
$\sigma$-derived winds with AGB feedback.  The metal enrichment of the
IGM at early times is significantly higher when the star formation
from smaller galaxies is included; therefore we use only higher
resolution models to explore the IGM enrichment at $z>2$.  Also, our
models suggest an overestimated SFRD at the lowest redshifts.  This
indicates that some other form of feedback is needed to suppress star
formation in massive galaxies at late times \citep[such as AGN
feedback; e.g.][]{cat07}.

The three models shown, the $\Phi$-derived winds and the
$\sigma$-derived winds with and without AGB feedback, are
indistinguishable in their global SFRD's above $z=4$ because the mass
loading factor and not wind speed is the largest determinant of star
formation~(OD06).  As explained in OD06, the earliest wind particles
have not had time to be re-accreted by galaxies, therefore the SFR is
regulated purely by how much mass is ejected ($\eta$).  The faster
wind velocities of the $\Phi$-derived model at $z<2$ decrease star
formation relative to the $\sigma$-derived model, because wind
particles are sent further away from galaxies making this gas harder
to re-accrete while also heating the IGM more and further curtailing
star formation.  We will quantify this recycling of winds in
\S\ref{sec: galstats}.

The average virial ratio of winds ($E_{\rm rat}$), defined as the ratio of
the kinetic energy to the potential at the launch position
($0.5\times\vw^2/-\Phi$), in the upper right panel shows how
$\sigma$-derived winds inject progressively less energy into their
surroundings with time. $E_{\rm rat}$ should be invariant across time for
$\Phi$-derived winds in its simplest form, but falls sharply below
$z=1$ due to our wind speed limits, and rises slightly at high-$z$ due
to the metal dependent $\vw$.  The faster $\Phi$-derived winds
spatially distribute more metals (solid lines in lower left panel) and
heat the IGM (lower right panel) to a greater extent than the
$\sigma$-derived winds.  We will show in \S\ref{sec: Zlocation} that
the cooler, less-enriched IGM of the $\sigma$-derived wind models
makes a significant difference in metals seen in quasar absorption
line spectra.  The gaseous metallicity (dashed lines in lower left
panel) is slightly higher in the $\Phi$-derived wind model despite
fewer overall stars formed, because fewer metals end up in stars.

Finally, we consider the impact of AGB feedback.  AGB stars do not add
any appreciable energy feedback, as demonstrated by the invariance
in the $E_{\rm rat}$ and volume-averaged temperature in the models with
(magenta lines) and without (blue) AGB feedback.  However, AGB stars
provide feedback in the form of returning gas mass to the ISM, which
is now available to create further generations of stars.  For instance,
with AGB feedback the SFRD at $z=0$ is increased by nearly a factor of
two compared to without.  Hence models that do not include such stellar
evolutionary processes may not be correctly predicting the SFRD history.

\section{The Universal Metal Budget} \label{sec: Zbudget}

In this section, we will investigate what the sources of cosmic metals
are, and where these metals end up.  To do so, we examine quantities
summed over our entire simulation volumes, with special attention to
the three different sources of metals: Type II SNe, Type Ia SNe, and
AGB stars.  We also discuss the impact of AGB feedback and
$\sigma$-derived winds.

\subsection{Sources of Metals} \label{sec: Zsource}

\subsubsection{The Stellar Baryonic Content} \label{sec: starcontent}

Since stars produce metals, we first examine the evolution of the stellar
baryonic content.  The key aspect for metal evolution is that stellar
recycling provides new fuel for star formation.  For the Chabrier IMF
assumed in our simulations, supernovae return $\sim20\%$ of stellar mass
instantaneously back into the ISM (which is double that for a Salpeter
IMF), and delayed feedback will eventually return another $\sim30\%$
over a Hubble timescale.  With half the gas being returned to the ISM,
most of it ($\sim80\%$) on timescales of less than a Gyr, subsequent
generations of stars can form, leading to greater metal enrichment.
The difference between $\sigma$-derived wind models with and without AGB
feedback in the SFRD plot (upper left panel of Figure~\ref{fig:global})
demonstrates how AGB feedback makes available more gas for star formation
at later times.

Despite more star formation in the l64n256vzw-$\sigma$ model with AGB
feedback, slightly more mass is in stars at $z=0$ in the no-AGB
feedback model by $z=0$ ($\Omega_{*}(z=0)/\Omega_{b}=0.081$ vs. 0.071)
since there is no mass loss from long-lived stars.  However, if we
count all stars (including short-lived stars undergoing SNe) formed
over the lifetimes of our simulated universes, the AGB feedback model
forms 37\% more stars ($\Omega_{*}(all)/\Omega_{b}=0.138$ vs. 0.101).
This is the more relevant quantity when considering the metal budget
of the local Universe.  Another way to think of this is that the stars
in today's local Universe account for
($\Omega_{*}(z=0)/\Omega_{*}(all)$=) 52\% of all the stars that have
ever existed, assuming the mass in short-lived stars is negligible (a
safe assumption in the low-activity local Universe).

\begin{figure} 
\includegraphics[scale=0.80]{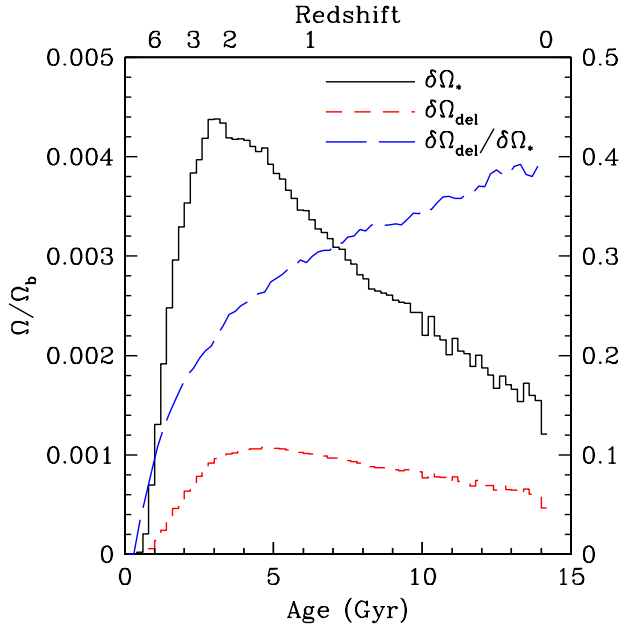}
\caption[]{Global quantities binned in 200 Myr bin of stars formed
($\delta\Omega_{*}$, solid black line) and mass lost via delayed
feedback ($\delta\Omega_{\rm del}$, dashed red line) relative to
$\Omega_b$ in the 32 $\hmpc$ $2\times256^3$ simulation.  The ratio of
the two is the long-dashed blue line.}
\label{fig:omegastar}
\end{figure}

Figure~\ref{fig:omegastar} shows the amount of baryons formed into
stars, $\delta\Omega_{*}$ (i.e. the star formation rate), and the amount
of stellar mass lost via delayed feedback, $\delta\Omega_{\rm del}$, as a
function of time.  The ratio $\delta\Omega_{\rm del}/\delta\Omega_{*}$
increases as more generations of stars are continuously formed with
each generation contributing to $\delta\Omega_{\rm del}$.  The quantity
$\delta\Omega_{\rm del}/\delta\Omega_{*}$ should approach 0.3 for a steady
star formation rate in a Hubble time, because 30\% of the stellar mass
is returned via delayed feedback, but this is actually exceeded since
the star formation is declining at late times.  The amount of material
lost via delayed feedback over the history of the Universe is found by
subtracting the stars at $z=0$ from the number of long-lived stars
formed over the history of the Universe.
\begin{equation}
\Omega_{\rm del} = \int_{0 {\rm Gyr}}^{14 {\rm Gyr}} \delta\Omega_{*}(t)(1-f_{\rm SN})dt - \Omega_{*}(z=0)
\end{equation}
where we assume $\Omega_{*}(z=0)$, the stars in today's Universe, has
a negligible quantity of short-lived stars.  The amount of delayed
feedback is $\Omega_{\rm del} = 0.039$ or 55\% of the stellar baryons in
today's Universe, which is a significant quantity.

\subsubsection{The Metal Content} \label{sec: Zcontent}

\begin{figure*}
\includegraphics[scale=0.80]{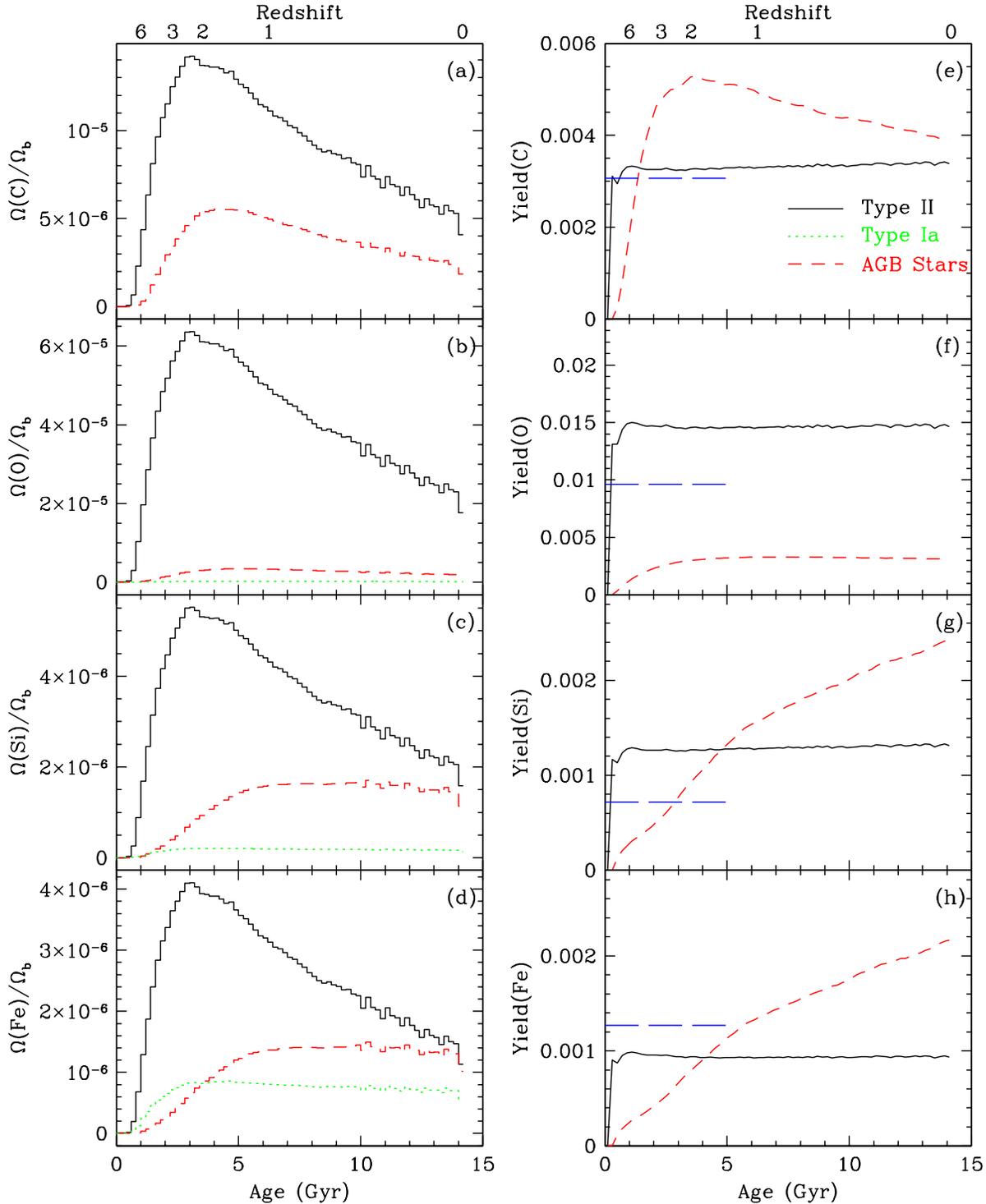}
\caption[]{Metal production by species relative to $\Omega_b$ along
with their yields plotted in 200 Myr bins in the 32 $\hmpc$
$2\times256^3$ simulation.  Panels (a-d) trace the amount of carbon,
oxygen, silicon, and iron respectively returned to the gas phase by
Type II SNe (black solid lines), Type Ia SNe (green dotted lines), and
AGB stars (red dashed lines).  The yields for the various species are
calculated in Panels (e-h) by taking lines in Panels (a-d) and
dividing by the matching line types in Figure~\ref{fig:omegastar};
long dashed lines on the left indicate \citet{and89} solar values.
Type II SNe yields remain relatively constant despite
metallicity-dependent yields.  The AGB carbon yields strongly depend
on $z$ and peak when 2-4 $M_{\odot}$ die.  AGB silicon and iron yields
are simply reprocessed metals formed in SNe and reflect the ensemble
metallicities of all stars undergoing AGB feedback at a given
redshift.}
\label{fig:yield}
\end{figure*}

We segue into our discussion of metals by plotting the global
production of the 4 species tracked by their source (Type II SNe, Type
Ia SNe, or AGB stars) in the left panels of Figure~\ref{fig:yield}.
Metals produced by Type II SNe depend on the current star formation
rate ($\delta\Omega_{*}$), which is apparent by the fact that the solid
lines in the left panels have roughly the similar shape to the global
stellar mass accumulation rate (Figure~\ref{fig:omegastar}).  Type II
SNe dominate the enrichment for all four elements at all redshifts,
hence global chemical enrichment is reasonably approximated by current
star formation alone.

Dividing the amount of each type of metal formed via Type II SNe by
$\Omega_{*}$ gives the SN yield of that element shown as solid lines
in the right panels of Figure~\ref{fig:yield}.  These yields should
and do match the Type~II SNe yield tables that are an input to the
simulations.  The yields do not evolve significantly because there is
only weak metallicity dependence in the \citet{lim05} yields employed
here.  Note that previous work has generally assumed no metallicity
dependence.  Above $z=6$, the yields are slightly lower due to less
metals injected by low-metallicity stars.  A slight upturn is
noticeable at lower redshift for carbon and silicon as their yields
increase with metallicity.

Turning to metals injected via AGB feedback (dashed red lines in
Figure~\ref{fig:yield}), we find more complex behavior that varies
among the species.  The left panels show that AGB feedback is an
important source of carbon ($\sim 30\%$ by $z=0$), iron and silicon
(both $\sim 25\%$ by $z=0$), but is a negligible contributor for
oxygen.  Dividing these values by $\Omega_{\rm del}$ results in the global
AGB yields as a function of redshift in the right panels.  For carbon
and oxygen, these are a summation of the yield values in our input
tables for all stars of various ages and metallicities undergoing AGB
feedback.

The carbon yields shows the most interesting evolution, which are
indicative of the underlying processes of stellar evolution in
different stars.  As explained in \S\ref{sec: AGBfeedback} and shown
in Figure~\ref{fig:cyield}, the most massive and short-lived AGB stars
(4-8 $M_{\odot}$) have hot-bottom burning that destroys carbon; these
are the stars losing mass via AGB feedback the most at very high-$z$.
Between 2-4 $M_{\odot}$, the third dredge-up makes AGB stars into
carbon stars with very high carbon yields from stars dying 200 Myr to
1 Gyr after their formation.  At $z\sim2$, carbon stars dominate the
carbon yields, but then less massive stars ($<2 M_{\odot}$) without
the third dredge-up begin to reach the AGB phase, and the ensemble AGB
carbon yield begins to decline.

Oxygen is burned in AGB stars, resulting in a net decrease in its
overall content as a result of delayed feedback.  Accounting for a
minor contribution from Type Ia SNe, the vast majority of gaseous
oxygen is synthesized in Type II SNe.  Hence oxygen is the ideal
species to trace the cosmic evolution of Type II SNe.

The AGB yields of silicon and iron may at first be surprising
considering that AGB stars do not process these elements.  These
yields reflect the ensemble metallicities of mass loss from AGB stars,
since they neither create nor destroy heavier elements at any
significant rate, but instead simply regurgitate them.  Most
surprising is that more iron is ejected from AGB stars than Type Ia
SNe.  The difference between these two forms of feedback associated
with stars is that the iron yield of Type Ia SNe is nearly a half
(i.e. 0.6 $M_{\odot}$ formed per 1.4 $M_{\odot}$ SNe) and should
significantly enrich its local environment, while the iron lost from
AGB stars should have a slightly lower yield than the surrounding gas
metallicity since these stars are older and hence less enriched.

It is curious and probably not correct that iron and silicon AGB
yields exceed solar metallicities by as much a factor of $2-3$ by
$z=0$.  This means that at $z=0$, the average AGB star is at least
$2-3\times Z_{\odot}$, which is almost definitely too high when stars
younger than the Sun in the Milky Way disk are $\sim Z_{\odot}$
\citep{twa80}.  Even though most $z=0$ AGB mass loss comes from stars
younger than the Sun since most AGB feedback occurs within 1 Gyr for a
Chabrier IMF, the extremely super-solar metallicities are indicative
of too much late star formation.  Reasons for this include too much
star formation in massive systems in our simulations, as well as our
slightly high value of $\Omega_b=0.048$ (instead of the currently more
canonical $\Omega_b=0.044$).

Table \ref{table:yields} summarizes the sources of metals at $z=2$
(roughly $10^{10}$ years ago) and $z=0$ for the l32n256vzw-$\sigma$
simulation.  These can be compared to available observational constraints.
Using oxygen, the global metallicity averaged over all baryons is
$\langle Z_b\rangle(z=2)\sim0.064 Z_{\odot}$ rising to $\langle
Z_b\rangle(z=0)\sim0.23 Z_{\odot}$.  These values are remarkably
similar to those derived by \citet{bou07} (hereafter B07) ($\langle
Z_b\rangle(z=2)\sim0.056$ and $\langle Z_b\rangle(z=0)\sim0.20 Z_{\odot}$)
where they assumed a Salpeter IMF-weighted metallicity yield of 0.024
and integrated over the star formation history of the Universe from
\citet{col01}.  While encouraging, this comparison is highly preliminary
owing to many systematics, such as the fact that our simulations produce
too many stars overall, and the assumption of a Salpeter IMF at all
times is probably not consistent with observations~\citep[see e.g.][and
discussions therein]{dav08a}.  It is hoped that improving observations
will enable more interesting constraints on cosmic chemical evolution
models.

\begin{table*}
\caption{Sources of Feedback}
\begin{tabular}{lccccccc}
\hline
Species & $\Omega/\Omega_b$ & \% Type II & \% Type Ia & \% AGB Stars & \% AGB Processing \\
\hline
$z = 2$ \\
\hline
Baryons & --         &  0.87 & .0016 &  0.83 &   --   \\
Carbon  & 1.677e-04  & 85.70 &   0.0$^a$ & 23.25 & +14.30 \\
Oxygen  & 6.123e-04  & 105.4 &  0.32 &  3.48 &  -5.75 \\
Silicon & 5.642e-05  & 96.50 &  3.57 &  9.30 &  +0.00 \\
Iron    & 5.013e-05  & 83.49 & 16.51 &  9.23 &  +0.00 \\
\hline
$z = 0$ \\
\hline 
Baryons & --         &  3.74 & .0098 &  5.42 &   --   \\
Carbon  & 6.566e-04  & 95.35 &   0.0$^a$ & 38.29 &  +4.65 \\
Oxygen  & 2.194e-03  & 126.1 &  0.52 &  7.68 & -26.60 \\
Silicon & 2.549e-04  & 95.29 &  4.72 & 33.89 &  +0.00 \\
Iron    & 2.265e-04  & 78.32 & 21.68 & 33.16 &  +0.00 \\
\hline 
\end{tabular}
\parbox{15 cm}{ $^a$These models were run with no Type Ia yields for
carbon.  These carbon yields are fractionally insignificant anyway.  }
\label{table:yields}
\end{table*}

While it is well-known that Type II SNe dominate carbon, oxygen, and
silicon production, it may be somewhat surprising for iron,
considering that Type Ia SNe are often assumed to be the primary
producers of iron; however, this is actually only true in environments
dominated by older stars.  Long-lived stars also destroy oxygen,
eliminating 20\% of the oxygen formed by Type II SNe.  Processing of
oxygen by AGB stars helps to move oxygen abundances from
alpha-enhanced levels to solar levels.  Of course, long-lived stars do
make a net surplus of both carbon and oxygen in post-Main Sequence
nucleosynthesis, however most of these metals remain locked in stellar
remnants, which we do not track in our simulations and are not
included in this table.  \cite{fuk04} estimate a $\langle
Z_b\rangle(z=0)\sim0.68 Z_{\odot}$ for all metals including those in
remnants, which exceeds the metals not locked up (i.e. the ones we
track) by a factor of a few.

Even though AGB feedback injects an appreciable amount of carbon into
surrounding gas ($\sim40$\% of carbon injected via Type II SNe), the
net surplus of carbon resulting from AGB feedback is only $\sim5$\% of
Type II SNe by $z=0$, because much of this carbon is coming from stars
with high metallicity already.  Carbon stars ($\sim2-4 M_{\odot}$) add
to the overall cosmic carbon abundance while higher and lower mass AGB
stars reduce the amount of carbon.  A larger impact on carbon
production comes from recycled gas that enables more Type II SNe; as
noted before, 37\% more stars form in simulations that include AGB
feedback.  Stars also lose mass via AGB feedback when they have moved
away from the sites of their formation, and can directly enrich
metal-poorer areas such as the ICM and intra-group medium.  Hence the
location of feedback from AGB stars and Type Ia SNe turns out to be
important for understanding enrichment in various environments.  We
consider this topic next.

\subsection{The Location of Metals} \label{sec: Zlocation}

B07 calculated from observations that metals migrated from gas to
stars between $z=2\rightarrow0$ as metals fall back into the deeper
potential wells of growing galaxies, and are more likely to remain
there as star formation-driven winds decline at low-$z$.  Our new
simulations generally agree with the results of B07 that about
one-third of metals are in stars at $z=2$, increasing to two-thirds by
$z=0$.

\begin{figure}
\includegraphics[scale=0.75]{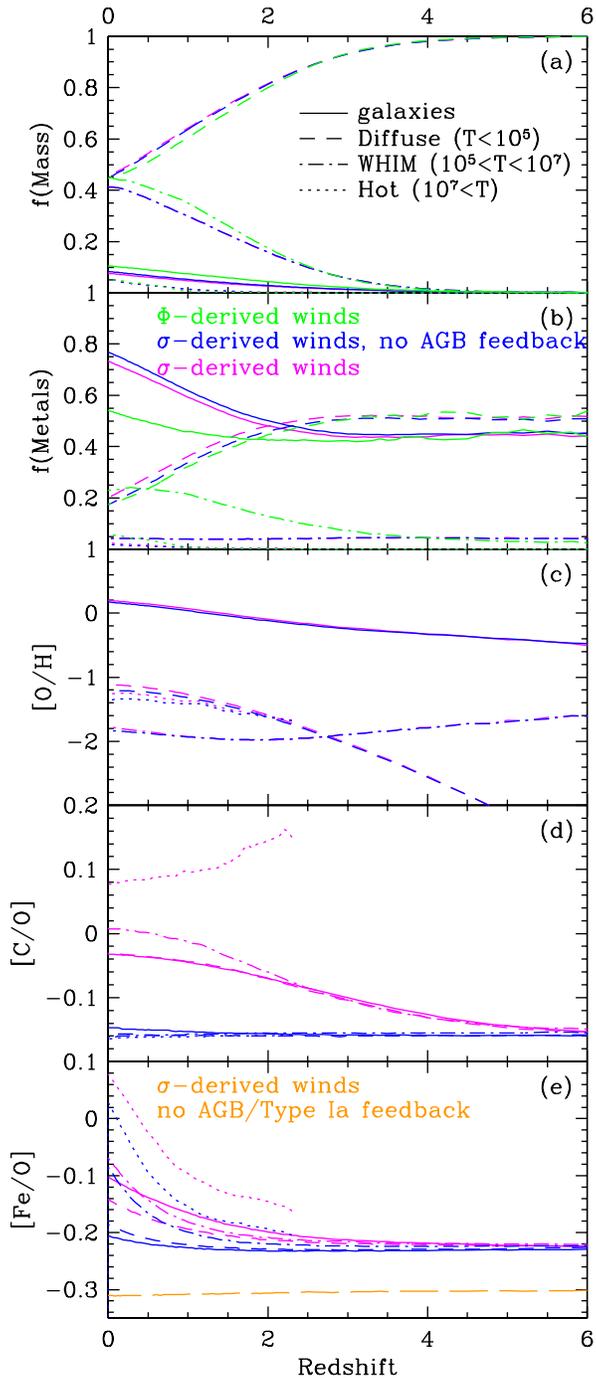}
\caption[]{The evolution of of mass and metals by baryonic phase in
the 64$\hmpc$, $2\times256^3$ simulations with and without AGB
feedback, and the 32$\hmpc$, $2\times128^3$ simulation with the old
potential-derived winds (shown only in top two panels).  The slower
$\sigma$-derived winds at low redshift do not inject nearly as many
metals into the WHIM.  The addition of AGB feedback does not change
the baryonic mass fractions (Panel a) and increases slightly the
amount of metals in galaxies (Panel b).  Oxygen metallicity (Panel c),
which traces Type II SNe, remains nearly unchanged, but delayed
feedback boosts carbon relative to oxygen (Panel d), and iron is
increases at late times relative to oxygen (Panel e) in hot gas mostly
due to Type Ia SNe.  The global [Fe/O] in a test simulation without
any delayed feedback is shown as the long dashed orange line in Panel
(e).  }
\label{fig:Zbudget}
\end{figure}

In Figure~\ref{fig:Zbudget} we subdivide mass and metals further by
their baryonic phases.  The top panel shows the baryonic fraction in
diffuse gas ($T<10^5$), warm-hot intergalactic medium (WHIM) gas
($10^5<T<10^7$), hot gas in clusters ($T>10^7$), and galaxies (stars
and ISM) for both the l64n256vzw-$\sigma$, l64n256vzw-$\sigma$-nagb,
l32n128vzw-$\Phi$ models as a function of redshift.  The steady
transfer of gas from diffuse to WHIM and hot IGM between
$z=3\rightarrow0$ was first noted by \citet{cen99} and \citet{dav99},
and our current results are similar to those from \citet{cen06a} using
nearly the same cosmology.  The addition of AGB feedback does not
appreciably change any of these values.

$\sigma$-derived winds have 5\% fewer baryons in the WHIM than
$\Phi$-derived winds at $z=0$ due to weaker winds at late times.
\citet{cen06a} find that their baryon fraction in the WHIM increases
by about 10\% owing to galactic superwinds, which matches our results
(not shown).  The metal fraction in the WHIM is basically constant in
the $\sigma$-derived winds, in stark contrast to the $\Phi$-derived
wind model (green lines; also see Figure~1 of DO07), which shows that
fraction of metals in WHIM grow steadily from $z=4\rightarrow 0$,
reaching 25\% today.  The much slower $\sigma$-derived wind velocities
at late epochs are mostly unable to shock-heat metals above
temperatures where metal-line cooling dominates, and instead the gas
cools efficiently.  Although the addition of AGN feedback is unlikely
to change the metal content of baryons by more than a few percent
(B07), the effect of $>1000$ $\kms$ winds from QSO's such as those
observed by \citet{tre07} could be appreciable for metals in the WHIM
and hot phases.

The total metal budget by baryonic phase in the second panel shows a
minor change owing to AGB feedback, namely that 5\% less metals are
found in the galaxies, with those metals instead being located in the
diffuse IGM.  This is because increased star formation from gas made
available via delayed feedback results in more winds that expel
metals.  

\subsubsection{Oxygen in the WHIM}

The third panel of Figure~\ref{fig:Zbudget} shows the oxygen
metallicity in various baryonic phases.  Overall, oxygen metallicities
([O/H]) remain nearly identical (within 0.1 dex) with the addition of
AGB feedback.  The 37\% increase in star formation and therefore
oxygen production from Type II SNe is counterbalanced by a 20\%
decrease due to the AGB processing of oxygen.  Galactic baryons show
slightly super-solar oxygen abundance, while the abundances in diffuse
and hot phases are nearly one-tenth solar.

The WHIM oxygen abundance is relatively constant with redshift, and
shows [O/H]=$-1.76$ at $z=0$.  Our simulations produce two distinct
types of WHIM: (1) the unenriched majority formed via the shock
heating resulting from structural growth, and (2) the WHIM formed by
feedback, which is significantly enriched.  The weaker
$\sigma$-derived winds form very little of the latter.  While
simulations suggest, by comparison with observed $\OVI$ absorbers,
that the WHIM metallicity should be around one-tenth
solar~\citep{cen01,che03}, it remains to be seen whether the
$\sigma$-derived winds are in conflict with $\OVI$ observations.
Since $\OVI$ arises in both photoionized and collisionally-ionized
gas, it could be that enough $\OVI$ is present in photoionized gas to
explain the observed number density of such systems.  Moreover,
non-equilibrium ionization effects could be important~\citep{cen06b}.
A careful comparison with $\OVI$ observations is planned, but is
beyond the scope of this paper.  

For now, we note that the oxygen abundance in the WHIM may be an
interesting probe of feedback strength.  \citet{som08} come to the
same conclusion at $z=3$ when tracing oxygen content by temperature,
noting that the stronger feedback by a top-heavy IMF produces
significantly greater amounts of oxygen in the WHIM.  Extremely fast
winds ($\vw>1000\kms$) emanating from AGN \citep{tre07} will also
create more enriched WHIM.

\subsubsection{Carbon in the IGM}

In the bottom two panels of Figure~\ref{fig:Zbudget} we show the
carbon and iron abundance relative to oxygen, a tracer of Type II SNe
enrichment, to emphasize the differences in these two species
influenced by delayed feedback.  For carbon especially, AGB feedback
has a significant impact.  [C/O] evolution shows an obvious increase
even at high-$z$, because the timescale for mass loss from carbon
stars is $0.2-1$~Gyr.  Every phase appears to evolve similarly with
their lines sometimes blending in Figure~\ref{fig:Zbudget} except the
hot phase, which has at least 50\% more carbon at $z>2$ .  Carbon
stars lose their mass near sites of star formation, and this carbon is
then blown out and shock-heated by a second generation of supernovae.
The net effect of AGB feedback on the $z=0$ carbon metallicity is
+0.22, +0.25, and +0.33 dex for diffuse, WHIM, and hot IGM
respectively.  This results in abundance ratios close to solar in all
phases.

\begin{figure}
\includegraphics[scale=0.90]{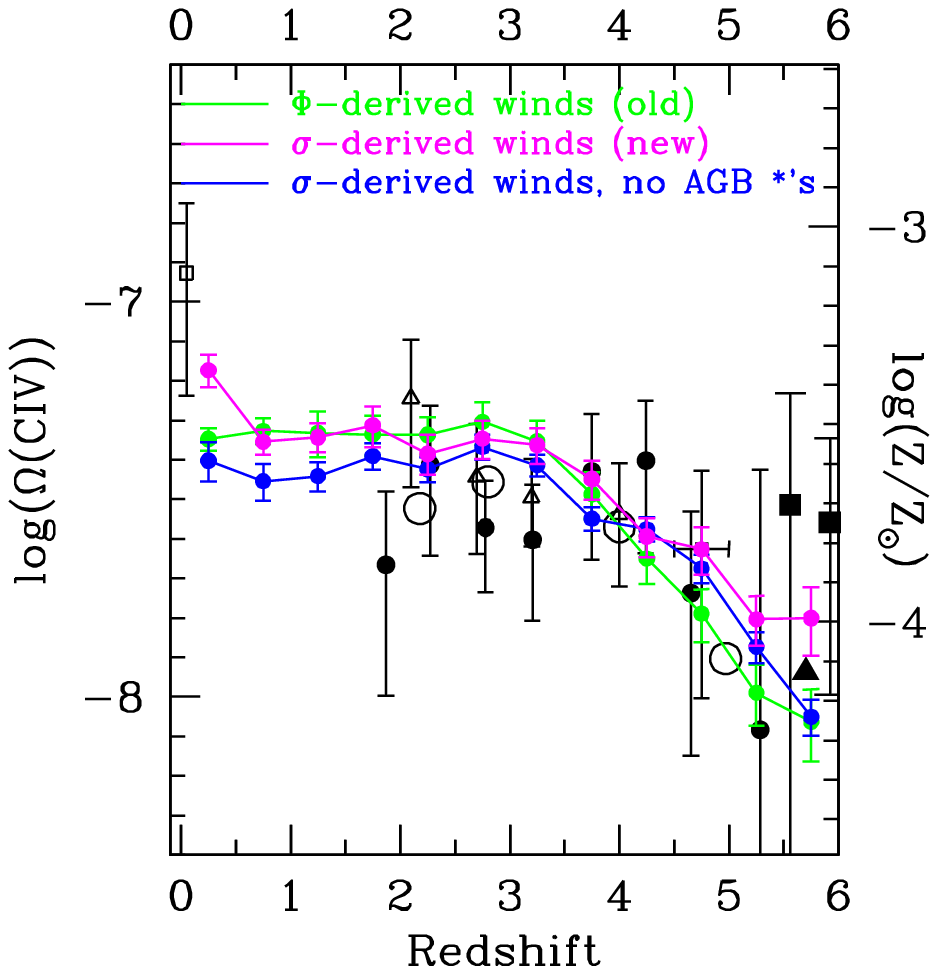}
\caption[]{Evolution of $\Omega$($\CIV$) from $z=6\rightarrow 0$ for
the $\Phi$-derived wind model with AGB feedback (green) and the
$\sigma$-derived wind model with and without AGB feedback (magenta and
blue).  Our models are compared to observations from \citet{son01}
(black circles), \citet{pet03} (small filled black squares),
\citet{bok03} (open triangles), \citet{fry03} (open square),
\citet{son05} (large open dots), \citet{sim06} (large black squares),
and \citet{rya06} (black triangle is a lower limit).  While all models
appear to fit the majority of the data, there are subtle differences
that we highlight between our models.  The new $\sigma$-derived winds
distribute metals more broadly boosting $\Omega$($\CIV$) at $z>5$,
while more carbon resulting from AGB feedback enriches the IGM in the
local Universe.  The three data points below $z=1.5$ are calculated
from 8 $\hmpc$ $2\times128^3$ simulations and from 16 $\hmpc$
$2\times256^3$ elsewhere, except in the case of the $\sigma$-derived
winds without AGB feedback in which case only the small box is used.
}
\label{fig:omegac4}
\end{figure}

A basic observational test of IGM enrichment models is the evolution
of $\Omega$($\CIV$), i.e. the mass density in $\CIV$ systems seen in
quasar absorption lines.  In Figure \ref{fig:omegac4} we plot
$\Omega$($\CIV$) from $z=6\rightarrow 0$ (see OD06 for exact method of
computing $\Omega$($\CIV$)) to see the effect of $\sigma$-derived
winds and AGB feedback.  In OD06 we reproduced the relative invariance
in the observed trend of $\Omega$($\CIV$) between $z=5\rightarrow 2$
by counterbalancing the increasing IGM carbon content by a similarly
lowering $\CIV$ ionization factor; the new $\sigma$-derived models
also match the observed trend quite well, for a similar reason.  The
addition of AGB feedback increases $\Omega$($\CIV$) by 70\% (+0.23
dex) at $z\la 0.5$, leading to a value consistent within the error
bars of $z\approx 0$ measurement by \citet{fry03}.

The main reason is that AGB feedback adds new fuel for star formation,
resulting in more $\CIV$ expelled into the IGM at late times.
Compared to the $\Phi$-derived winds, the $\sigma$-derived winds push
out more metals early better matching the high-$z$ $\CIV$ observations
of \citet{rya06} and \citet{sim06}.  The faster $\Phi$-derived winds
at low-$z$ raise the temperature of the metal-enriched IGM while
pushing the metals to lower overdensities and lowering the $\CIV$
ionization fraction.

The $\sigma$-derived wind model with AGB feedback is the first model
we have explored that is able to fit the entire range of
$\Omega$($\CIV$) observations from $z\sim 6\rightarrow 0$.  

The $\sigma$-derived wind model with AGB feedback achieves higher
$\Omega$($\CIV$) at $z>5$ and $z<1$ that at face value improves
agreement with observations.  While these data are uncertain and hence
one should not over-interpret this improved agreement, the main point
of this exercise is to show how our newly incorporated physical
processes could have observational consequences.  Furthermore, the
results are to be taken with caution, owing to the small box size of
our simulations (8 $\hmpc$ $2\times128^3$) below $z=1.5$; this volume
is not nearly large enough to form the large-scale structures in the
local Universe.  It is encouraging that the values derived from this
small box agree within the error bars with the larger
l16n256vzw-$\sigma$ simulations above $z=1.5$.  Future observations at
low-$z$ by the Cosmic Origins Spectrograph, and at high-$z$ with
advances in near-IR spectroscopy will allow more relevant and detailed
comparisons.

\subsubsection{Iron in the ICM}

The [Fe/O] evolution (bottom panel of Figure~\ref{fig:Zbudget}) is
dominated by Type II SNe until $z\sim 2$, at which point delayed
feedback processes of Type Ia SNe and AGB stars become important.  The
instantaneous component of the Type Ia SNe adds 19\% of the Type II
SNe iron yield, which raises the [Fe/O] everywhere from -0.30 to -0.23
(compare to global [Fe/O] in a simulation without any Type Ia's or AGB
feedback\footnote{We ran a l32n128vzw-$\sigma$ test simulation with only
Type II feedback, and the only major difference is the lack of iron
produced via Type Ia's relative to l32n128vzw-$\sigma$-noagb.}).  The
delayed component of Type Ia SNe adds only 9\% more of the Type II SNe
iron content generated over the lifetime of the Universe.  However,
the combination of the high iron yield (0.43) and the location of
enrichment often being low-metallicity regions away from the sites of
star formation meas delayed Type Ia's can have a significant
observational signature in a low-density medium such as the ICM.  The
net increase of all Type Ia SNe (delayed and instantaneous) on the
$z=0$ iron content is +0.32 dex in the hot component and +0.21 dex in
the WHIM.  AGB feedback increases iron content by allowing 37\% more
stars to form; this extra iron primarily remains in galaxies, but
increases iron in the hot component by +0.15 dex and the WHIM by +0.07
dex.  Overall, the hot iron metallicity increases by nearly 3$\times$
with the addition of Type Ia and AGB feedback.

To demonstrate the effect on observables, we calculate the free-free
emission-weighted [Fe/H] of the ICM in clusters/groups with
temperatures in excess of 0.316 keV at $z=0$ in the l64n256vzw models
to simulate the X-ray observations.  We identify large bound systems
in the simulations using a spherical overdensity
algorithm~\citep[see][for description]{fin06}.  The average of over
130 clusters/groups is -1.11 without any delayed feedback, -0.78 with
Type Ia SNe included, and -0.57 with AGB feedback also included.
Hence the addition of delayed forms of feedback increases the ICM
[Fe/H] by $3.4\times$.  This is now in the range for the canonical ICM
metallicities of around 0.3~solar, as well as for groups between 0.316
and 3.16~keV as observed by \citet{hel00} (they found
$-0.60<$[Fe/H]$<-0.36$).  Of course, X-ray emission at $\sim 1$~keV
has a significant contribution from metal lines, and observations can
be subject to surface brightness effects~\citep[e.g.][]{mul00}, so
this comparison is only preliminary.  A more thorough comparison of
simulations to ICM X-ray observations is in preparation \citep{dav08b}.

To summarize this section, we showed that Type II SNe remain the
dominant mode of global production for each species we track, usually
by a large margin.  When considering metals not locked up in stellar
remnants (i.e. observable metals), Type Ia SNe only produce 22\% of
the cosmic iron and AGB stars only contribute 5\% to the cosmic carbon
abundance.  Mass feedback from long-lived stars allows metals to be
recycled and form new generations of stars, increasing late-time star
formation by $\sim 30-40\%$.  More importantly, the location of metals
injected by delayed modes of feedback can significantly impact
metallicities in specific environments.  The IGM carbon content,
probably the best current tracer of IGM metallicity, increases by 70\%
by $z=0$ when AGB feedback is included.  The iron content observed in
the ICM at least triples, primarily due to Type Ia SNe.  Delayed
metallicity enrichment appears to heavily affect the enrichment
patterns of the low density gas of the IGM and ICM where there are
relative few metals, compared to galaxies where we find the
metallicity signatures of Type II SNe dominate.  Although our
simulations do not produce large passive systems at the present epoch,
it is likely that delayed modes of feedback will be important for
setting the metallicity in and around such systems as well.  Hence
incorporating delayed feedback is necessary for properly understanding
how metals trace star formation in many well-studied environments.



\section{Galaxies and Feedback} \label{sec: galstats}

Thus far we have examined energy balance and metallicity budget from a
global perspective.  In this section we investigate such issues from
the perspective of individual galaxies.  We will answer such questions
as: What galaxies are dominating each type of feedback (mass, energy,
and metallicity)?  How does this evolves with redshift?  Do winds
actually leave galaxy haloes and reach the IGM?  What types of galaxies
are enriching the IGM at various epochs?  The key concept from this
section is wind recycling; i.e. the products of feedback do not remain
in the IGM, but instead are either constantly cycled between the IGM
and galaxies or never escape their parent haloes in the first place,
and are better described as halo fountains.

During each simulation, all particles entering a wind 
are output to a file.  The originating
galaxy is identified, and the eventual reaccretion into star-forming
gas is tracked.  In this way wind recycling can be quantified in
galaxy mass and environment.  Throughout this section we will use
SKID-derived galaxy masses, which match our on-the-fly FOF galaxy
finder for the vast majority of cases (cf. \S\ref{sec: groupfinder}).
We use our favored $\sigma$-derived wind simulations in our following
analysis, unless otherwise mentioned.

\subsection{Feedback as a Function of Galaxy Mass} \label{sec: galmass}

\begin{figure*}
\includegraphics[scale=0.80]{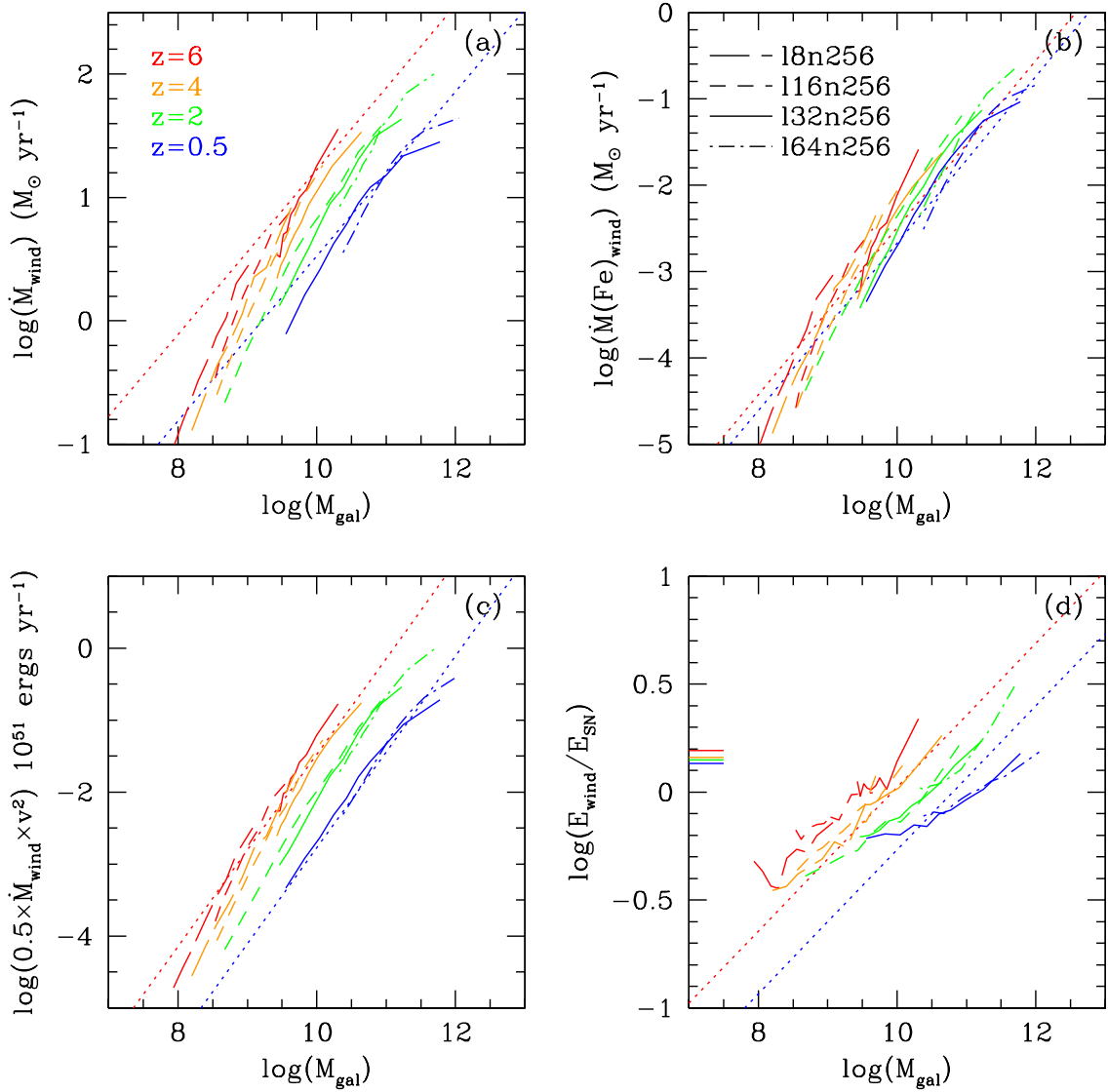}
\caption[]{ Feedback properties are shown as a function of galaxy mass
(derived by SKID) for 4 redshifts for our $\sigma$-derived wind
simulations at various box sizes.  Red and blue dotted lines are our
``toy models'' calculated using momentum-driven wind relations and
assuming specific star formation rates of 0.1 and 1.0 Gyr$^{-1}$, and
metallicities of $Z=0.3 Z_{\odot}$ and 1.0 for $z=6$ and 0.5
respectively.  Mass and metal feedback (Panels (a) and (b)) are
predicted to go as $M_{\rm gal}^{2/3}$, although metallicity feedback
additionally depends on the gas mass-metallicity of galaxies, where
$Z\propto M_{\rm gal}^{0.3}$.  Energy feedback, in $10^{51}$ erg units (Panel
(c)) follows the predicted $M_{\rm gal}^{4/3}$ relation very closely, and
the ratio wind energy efficiency, $E_{\rm wind}/E_{\rm SN}$, (Panel (d)) rises
nearly as $M_{\rm gal}^{1/3}$ for most redshifts.  The lines on the left
of Panel (d) show the average $E_{\rm wind}/E_{\rm SN}$ efficiency in the
l32n256 box is declining slightly, although these values may be lower
when lower mass galaxies not resolved in this box are included. }
\label{fig:mfunc}
\end{figure*}

Figure~\ref{fig:mfunc} quantifies mass (upper left), metal (upper
right) and energy (lower panels) feedback, as a function of galaxy
mass in our $\sigma$-derived wind simulations at four chosen redshifts
($z=6,4,2,0.5$).  We choose $z=0.5$ to represent the local Universe
rather than $z=0$, because we want to follow the evolution of wind
materials after they are launched and consider 5 Gyr a compromise as
enough time for the winds cycle to play out, but not too much such
that the cosmological evolution is overly significant.  The upper left
panel shows the mass loss rate in outflows as a function of galaxy
baryonic mass.  At a given galaxy mass, the outflow rate goes down
with time, by roughly a factor of 10 from $z=6\rightarrow 0.5$.
Remember that this is the rate of mass being driven from the galaxy's
star-forming region; whether the material makes it to the IGM or
remains trapped within the galactic halo will be examined later.

Along with the results of our simulations, we plot two simple "toy
models" of feedback behavior corresponding to galaxies forming stars
at constant specific star formation rates (i.e. star formation rate
per unit stellar mass) of 1.0 and 0.1 Gyr$^{-1}$; these roughly
correspond to typical star forming galaxies at $z=6$ and $z=0.5$
respectively.  The momentum-driven wind model predicts that mass
feedback should go as $\dot{M}_{\rm wind}\propto$ SFR$/\sigma \propto$
SFR$\times M_{\rm gal}^{-1/3} H(z)^{-1/3}$.  Making the reasonable
assumption that SFR$\propto M_{\rm gal}$ as typically found in
simulations~\citep[e.g.][]{dav08a}, then this simple model would
predict $\dot{M}_{\rm wind}\propto M_{\rm gal}^{2/3} H(z)^{-1/3}$.
The dotted lines in Figure \ref{fig:mfunc} show these relations for
our toy models.  The red dotted line fits well to $z=6$ at $M_{\rm
gal}=10^{9.5-10} M_{\odot}$ showing these galaxies efficiently
doubling their mass every 1.0 Gyr, while the doubling time is around
10 Gyr for galaxies at $z=0.5$.

The typical mass outflow rate reduces with time for two reasons:
First, the star formation rates are lower owing to lower accretion
rates from the IGM, as discussed in SH03b and OD06 and as observed by
\citet{per05, cap06, pap06}.  Second, galaxies grow larger with time
and the mass loading factors drop; this even despite the $H(z)^{-1/3}$
factor that actually increases $\eta$ for a galaxy of the same mass at
lower redshift.  Hence the outflow rates qualitatively follow the
trend seen in observations that at high redshifts, outflows are
ubiquitous and strong, while at the present epoch it is rare to find
galaxies that are expelling significant amounts of mass.


Figure~\ref{fig:mfunc}, upper right panel, shows the mass of metals
launched as wind particles, which is $\dot{M}(Z)_{\rm wind}\propto
M_{\rm gal}^{2/3} H(z)^{-1/3} Z_{\rm gal}$.  For concreteness we follow the
iron mass, although other species show similar trends; recall that
even at $z=0$ 93\% of iron is produced in Type II SNe, and the
fraction is higher at higher redshifts.  This relation can be thought
of as the $M_{\rm gal}-\dot{M}_{\rm wind}$ relation shown in the upper left
panel convolved with the star formation-weighted gas mass-metallicity
relations of galaxies at the chosen redshifts, since it is this gas
that is being driven out in outflows.  \citet{fin08} showed how our
{\it 'vzw'} model reproduces the slope and scatter of the
mass-metallicity relationship of galaxies as observed by \citet{erb06}
for $z=2$ Lyman break galaxies.  Again, we plot two dotted lines
corresponding to our toy models with a $Z\propto M_{\rm gal}^{0.3}$
dependence accounting for the mass-metallicity relationship normalized
to $z=0.3$ and $1.0 Z_{\odot}$ at $10^{11} M_{\odot}$ for $z=6.0$ and
0.5 respectively.  The two toys models show little evolution (0.18 dex
decline from $z=6\rightarrow 0.5$, because the declining
$\dot{M}_{\rm wind}$ is counter-balanced by an increasing metallicity for
a given mass galaxy toward lower redshift.  In the l32n256vzw-$\sigma$
simulation, a $10^{10} M_{\odot} yr^{-1}$ galaxy injects
$7.7\times10^{-3} M_{\odot} yr^{-1}$ of iron at $z=6$ and
$2.2\times10^{-3} M_{\odot} yr^{-1}$ at $z=0.5$.

In the lower left panel we plot the total feedback energy per time
(i.e. feedback power) imparted into wind particles as a function of
galaxy mass.  This power is $\dot{E}_{\rm wind}=0.5\dot{M}_{\rm
wind}\times \vw^2\propto$ SFR$\times\sigma H(z)^{1/3}\propto M_{\rm
gal}^{4/3} H(z)^{1/3}$, using the same assumption of SFR $\propto
M_{\rm gal}$.  The feedback power is shown in units of $10^{51}$ ergs
yr$^{-1}$, which can be thought of as the number of SNe per year.  Our
simulations follow the trend of the toy models in terms of redshift
evolution, and show an even tighter agreement at the low mass end
versus $M_{\rm gal}$ than the mass feedback; the metallicity
dependence gives greater energies to winds from lower mass, less
metal-rich galaxies.  Energy feedback is an even stronger function of
galaxy mass than mass or metallicity feedback.

The bottom right panel in Figure \ref{fig:mfunc} shows feedback energy
relative to supernova energy ($E_{\rm wind}/E_{\rm SN}$).  This quantity
decreases with time, and increases with galaxy mass.  The toy models
shown represent $E_{\rm SN}\propto$ SFR $\propto M_{\rm gal}$, or
$E_{\rm wind}/E_{\rm SN}\propto M_{\rm gal}^{1/3} H(z)^{1/3}$.  At the low-mass
end, the simulations rise above the toy model and show more scatter
due to uncertainties in the the star formation rates of the smallest
galaxies, but the slope at most redshifts is correct.  At the
high-mass end, momentum-driven wind energy exceed the supernova
energy, which is physically allowed since the UV photons produced
during the entire lifetimes of massive stars drive winds in this
scenario (see OD06, Figure~4).  Summing this ratio globally over all
galaxies at each redshift, we obtain the values shown by the tick
marks on the left side of the panel.  Globally, the average
$E_{\rm wind}/E_{\rm SN}$ ratio exceeds unity at all redshifts (being around
1.2), and is surprisingly constant, declining less than 0.1 dex from
$z=6\rightarrow0.5$ in the new $\sigma$-derived formulation of the
winds.  The decline of wind energy feedback for a given mass galaxy
toward lower redshift is mostly counterbalanced by more massive
galaxies driving more energetic winds at these redshifts.  Less
massive galaxies unresolved in the l32n256 box are likely to lower
this value somewhat, so we only want to conclude that the wind energy
is similar to the supernova energy and stays remarkably unchanged with
redshift in the $\sigma$-derived momentum driven wind model.

To summarize, the momentum-driven wind simulations follow trends expected
from the input momentum-driven outflow scalings.  This is of course not
surprising, and at one level this is merely a consistency check that
the new wind prescription and the group finder are working correctly.
But this also gives some intuition regarding outflow properties as a
function of galaxy mass required to achieve the successes enjoyed by the
momentum-driven wind scenario.  For example, mass outflow rates should
correlate with galaxy mass, and outflow energy in typical galaxies is
comparable to, and perhaps exceeds, the total available supernova energy.
These trends provide constraints on wind driving mechanisms and inputs
to heuristic galaxy formation models such as semi-analytic models.

\subsection{Feedback by Volume} \label{sec: volume}

\begin{figure*}
\includegraphics[angle=-90,scale=0.66]{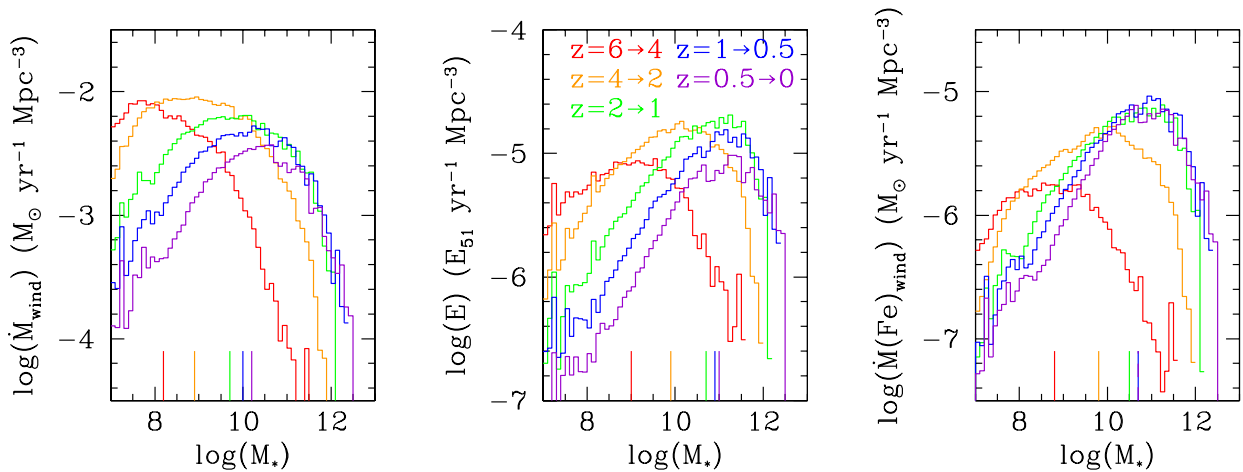}
\caption[]{Mass, metallicity, and energy feedback per cubic Mpc binned
by stellar mass (0.1 dex bins, derived by SKID) in 5 redshift bins
covering $z=6\rightarrow0$.  Histograms include data from 8, 16, 32,
and 64 $\hmpc$ in order to resolve a large range (5 decades) of galaxy
masses.  Vertical lines with colors corresponding to their redshift
range show the median galaxy producing each type of feedback.  For all
forms of feedback, the median $M_*$ increases by a factor of 100
between $z=6\rightarrow0$; most of this increase ($30\times$) is
between $z=6\rightarrow2$.  However each form of feedback favors a
different mass-scale: $10^{10.2} M_{\odot}$ for mass, $10^{11.0}
M_{\odot}$ for energy, and $10^{10.7} M_{\odot}$ for metallicity.  The
mass resolution limits are $1.9\times10^7 M_{\odot}$ above $z=4$ and
$1.5\times10^8 M_{\odot}$ below.  }
\label{fig:starfunc}
\end{figure*}

We shift from examining feedback trends in individual galaxies to studying
feedback trends per unit volume.  In order to facilitate observational
comparisons, we use use stellar mass, $M_*$, rather than baryonic mass.

In Figure \ref{fig:starfunc} we plot histograms binned in 0.1 dex
intervals of the three forms of feedback at five redshift bins from
$z=6\rightarrow0$ parameterized by the amount of feedback per cubic
Mpc.  These histograms include all SKID-identified galaxies in the
8, 16, 32, and 64 $\hmpc$ boxes in order to obtain the large dynamical
range covering over 5 decades of galaxy masses.  The less
computationally expensive l8n128vzw simulation was included in the
histograms between $z=0-1.5$ to probe the least massive galaxies in
this range, since the l16n256vzw run at the same resolution ends at
$z=1.5$.  We also plot the median $M_*$ of a galaxy contributing to
each type of feedback shown as vertical lines at the bottom of each
panel.

The mass, energy, and metal outflow rates peak at increasingly higher
galaxy masses with time.  The galaxy mass resolution limits are
$1.9\times10^7 M_{\odot}$ at $z\geq 4$ and $1.5\times10^8 M_{\odot}$
below; the peaks are mostly comfortably above these resolution limits,
indicating that we have the necessary resolution to resolve the source
of feedback across our simulation boxes for the three forms of
feedback.  The exceptions are the mass outflow rates at $z>2$, which
peak less than 1~dex from these limits.  The changing resolution limit
at $z=4$ raises concern whether the evolution above and below this
limit is real; however, the larger amount of evolution between
$z=4\rightarrow1$, despite an unchanging resolution limit shows that
evolution at $z<4$ is not just a resolution effect.

The median galaxy $M_*$ expelling gas increases by $\sim 100\times$
between $z=6\rightarrow0$, for all three forms of feedback.  The vast
majority of this growth occurs between $z=6\rightarrow2$, where the
median stellar mass increases by $\sim30\times$ in just 2.3 Gyr.  This
is the epoch of peak star formation in the Universe, so it is not
surprising that galaxies show the most growth in their stellar masses
then.  The baryonic mass (stars plus gas, not shown) also jumps
significantly, $\sim10-15\times$, but early small galaxies are more
gas rich making the jump less extreme.  Nevertheless, the evolution is
much slower in the 10 Gyr between $z=2\rightarrow0$ as both median
$M_*$ and $M_{\rm gal}$ at most triple and usually double in this
longer timespan.  This late growth could be an overestimate because of
our overestimated star formation rates at late times in the most
massive galaxies.  It is quite possible that if our simulations could
properly curtail massive galaxies from forming stars, median $M_*$
would fall toward $z=0$.

The same growth patterns for star formation-driven feedback are also
seen in the median galaxy weighted by SFR, which itself is highly
correlated with $M_{\rm gal}$ as \citet{dav08a} shows using these same
simulations.  Therefore the changing mass scales of star
formation-driven feedback reflect the hierarchical growth of galaxies
between $z=6\rightarrow0$, with most growth occurring before $z=2$.

While it is no coincidence that all three forms of feedback show
similar growth rates in median $M_*$ (and $M_{\rm gal}$), each
feedback form has a different mass preference and peaks at a different
epoch.  Mass feedback preferentially traces smaller galaxies due to
the inverse relationship of $\eta$ with $M_{\rm gal}$.  High mass
loading factors from small galaxies, which dominate star formation at
high redshift are necessary to curtail the star formation density at
early times (OD06) and fit galaxy luminosity functions at $z>3$
\citep{fin06, bouw07}.  The mass feedback density peaks at $z\sim3.5$
at $0.28 M_{\odot} yr^{-1} {\rm Mpc}^{-3}$.  The median galaxy in the
relatively nearby Universe bin ($z=0.5\rightarrow0$) has
$M_{*}=10^{10.2} M_{\odot}$, or 1/6th the \citet{bel03} $M^*$ value of
$10^{11.0} M_{\odot}$ in stellar mass.

The median $M_*$ for energy feedback is biased toward larger galaxies
for momentum-driven winds.  Our simulations suggest that the median
galaxy adding energy to the IGM is an $M^*$ galaxy ($M_*=10^{11}
M_{\odot}$) in the local Universe.  Observations disagree with this
prediction, since the prototypical local galaxy exhibiting feedback is
more like M82, an $\sim0.1 M^*$ galaxy.  Moreover most super-$M^*$
galaxies are red and dead, unable to generate star formation-driven
winds.  This again suggests that a more realistic truncation of star
formation in low-$z$ massive galaxies could alter the exact values
quoted here.  However, at high-$z$ where the typical galaxy appears to
be driving an outflow~\citep{erb06}, these results should be more
robust.  The energy feedback density peaks at $z\sim2.5$ at an
equivalent energy of $5\times10^{-4}$ SN yr$^{-1}$ Mpc$^{-3}$
($10^{51}$ ergs each).


The metallicity feedback grows the fastest of all forms of feedback
during the epoch of peak star formation, and does not obtain its
maximum density until slightly past $z=1$ ($\dot{M}(Fe)_{\rm
wind}=1.7\times10^{-4} M_{\odot}$ yr$^{-1}$ Mpc$^{-3}$).  The
metallicity feedback is simply the mass feedback modified by the
mass-metallicity relationship of galaxies, which favors higher $M_*$
and lower redshift.  The median $M_*$ at $z=0$ is $0.5 M^*$, or
$\sim5\times$ higher than the $0.1 L^*$ median galaxy as determined by
B07.  However, we do not think that this is an inconsistency due to an
overestimate in massive galaxy star formation at low redshift, but
instead a result of measuring different quantities.  We are measuring
the amount of feedback leaving a galaxy's star forming region whereas
B07 measures the feedback expelled from the galaxy reaching the IGM by
determining the effective yield as a function of $v_{\rm rot}$ .  As
we will show next, much of the mass and metals that are expelled never
leave their parent haloes and are reaccreted in a timescale often
much less than the Hubble time.  The leads us to introduce the
important concept of wind recycling.

\subsection{Wind Recycling} \label{sec: recycle}

What happens to the mass and metals once they are expelled from the
galaxies' star forming regions?  Is all feedback best described as
galactic superwinds or does some feedback never really escape from its
parent halo and should more accurately be considered a halo fountain?
To answer such questions we introduce the concept of wind recycling,
which plays an important role in feedback over cosmic time.  

By following wind particles by their particle IDs during the
simulation run, we can track how many times the same SPH particle is
recycled in a wind.  For the l32n256vzw-$\sigma$ simulation evolved
all the way to $z=0$ we find that a wind particle is launched an
average of 2.5 times; while 18.3\% of SPH particles are ever launched
in a wind, the summed number of wind launches equals 45.7\% of the
total number of SPH particles.  Wind recycling dominates over winds
being launched from galaxies for the first time-- the average wind
particle across time is more likely to have already been launched in a
wind!  The most important aspect is that wind material, once launched,
cannot be assumed to be lost from the galaxy forever and remaining in
the IGM.  This is true despite the fact that, in our momentum-driven
wind prescription, outflows are always ejected at speeds exceeding the
escape velocity of its parent galaxy.  Gravitational infall and
hydrodynamic effects both conspire to slow down outflows and
facilitate wind recycling.

\begin{figure}
\includegraphics[scale=0.80]{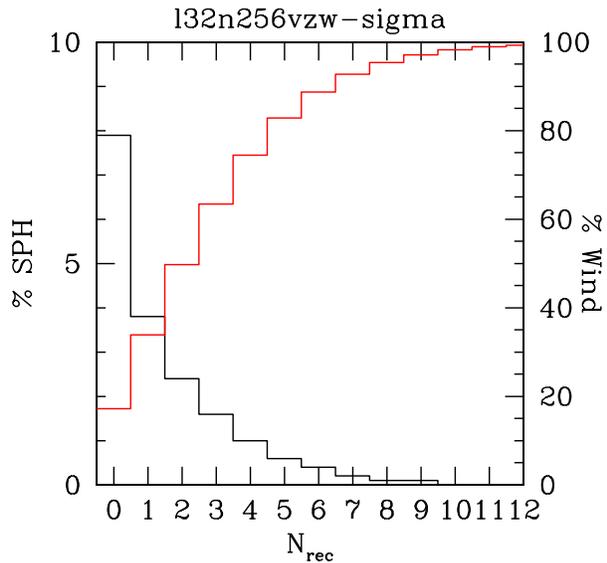}
\caption[]{The percentage of all SPH particles recycled $N_{\rm rec}$
 times by $z=0$ in the $32\hmpc$ $2\times256^3$ simulation is shown by
 the black histogram (left scale). $N_{\rm rec}=0$ means a wind particle
 is launched only once. The red line shows the cumulative sum of wind
 particles launched per bin (right scale, normalized to 100\%) and is
 the sum of $(N_{\rm rec}+1)$ multiplied by the black histogram added left
 to right.  50\% of winds are particles that have been or will be
 recycled 3 or more times indicating that wind recycling plays a
 significant if not dominant role in feedback.  }
\label{fig:windrec}
\end{figure}

Figure \ref{fig:windrec} displays histograms of the number of times
the same wind particle is recycled, $N_{\rm rec}$, in the l32n256
simulation.  Only 17\% of all winds are particles ejected one time and
therefore never recycled (i.e. $N_{\rm rec}=0$); this corresponds to 7.9\%
of all SPH particles.  The record-holder is a particle recycled an
astonishing 30 times indicating that probably in this case the term
halo fountain may be more appropriate than galactic superwind.
Perhaps the most telling statistic is half of the wind particles have
been recycled 3 or more times.  The continuous range in the number of
recycling times blurs the distinction between a galactic superwind and
a halo fountain, and suggests instead there is a continuum.

The concept of recycling is not unexpected, and has been predicted by
\citep{ber07} from semi-analytical models.  DO07 showed that metals
move from mean cosmic density at $z=2$ to an overdensity of 100 by
$z=0$ as baryons migrate into larger structures as part of cosmic
structural growth.  This means that most of the metals in the diffuse
IGM at $z\sim 2$ are in galactic haloes at $z\sim 0$.  Metals blown
out from early galaxies to low overdensities are later reaccreted in
the formation of larger structures, and blown out again.  Still, the
commonality of wind particles being recycled is surprising, especially
since momentum-driven winds are almost always ejected at velocities
well in excess of the escape velocity of the galaxy.  It is in fact
more appropriate to talk about how long it takes a wind particle to be
recycled instead of whether it will be recycled.  Metals injected by
galactic superwinds cannot be assumed to remain permanently in the
IGM; metals continuously cycle between galaxies and the IGM.  In an
upcoming paper we will quantify the ages of metals observed at
different absorption lines tracing different regions of the IGM.  For
now we note that the average ages of the metals in the IGM are
typically much shorter than the age of the Universe.

Do wind particles generally return to their parent galaxy, or do they
jump from galaxy to galaxy?  The answer is the former; in the vast
majority of cases a wind particle returns to either its parent galaxy
or the result of a merger involving the parent galaxy.  95\% of
recycled wind particles are re-launched from a galaxy of similar or
more mass than the previous recycling.  Of course galaxies grow anyway
under the hierarchical growth scenario, so a wind particle could join
a different galaxy that has itself grown larger than its parent.  By
considering wind particles recycled within 10\% of the Hubble time, we
can diminish the bias of galaxy growth and explore whether winds are
traveling from massive central galaxies to surrounding satellite
galaxies possibly affecting their dynamics and enrichment histories.
With this time limit, 97\% of winds return to more massive galaxies
indicating this is a less likely trend.  The number jumps to 99\% when
considering winds launched from $M_{\rm gal}>10^{11} M_{\odot}$,
indicating that it is harder to escape from a more massive galaxy.
Furthermore, a wind particle is rarely relaunched more than 1 comoving
$\hmpc$ apart in a 16 $\hmpc$ test simulation; the minority case
usually involves wind particles following the position of a fast
moving parent galaxy.  Winds rarely escape forever their parent halo,
and winds from large central galaxies do not appear to disrupt
satellite galaxies.

Another way to quantify wind recycling is using the summed amount of
material injected into winds.  We introduce the quantity
$\Omega_{\rm wind}$, which is the mass injected in winds, including
recycled gas.  Owing to recycling, there is no limit to how large this
can be.  Using our l32n256 run, we find that
$\Omega_{\rm wind}=0.39\Omega_{b}$-- i.e. an equivalent of 39\% of the
baryonic mass has been blown out in a galactic superwind by
$z=0$.\footnote{Although the total number of wind launches equals
45.7\% of all SPH particles, the typical wind particle is less massive
than the average SPH particle, because one half of a full SPH particle
remains when a star particle is spawned, and wind particles are more
likely to arise from these remaining SPH particles.}  This is
significantly larger than $\Omega_{*}=0.097$ due to mass loading
factors greater than one as well as recycling.

The other $\sigma$-derived wind runs to $z=0$ give different values:
$\Omega_{\rm wind}=0.59\Omega_b$ for the 8 $\hmpc$ box with $128^3$ SPH
particles, and $\Omega_{\rm wind}=0.23\Omega_b$ for the 32 $\hmpc$ box
also with $128^3$ particles.  $\Omega_{\rm wind}$ increases by 160\% when
mass resolution changes by $64\times$.  This jump is primarily due to
a 70\% increase in $\Omega_{*}$, but still leaves another 50\% likely
due the varying treatment of recycling at different resolutions.  This
lack of resolution convergence raises the question of whether the
amount of recycling is set by numerics, perhaps arising from the poor
treatment of SPH trying to model outflow processes occurring at
resolutions below the limit of our simulations.

The discontinuity in recycling rates between resolutions stems from
how long and how far a particle reaches once it has been launched.
Wind particles in the 32 $\hmpc$ $2\times256^3$ box with typical
masses of $3.2\times10^7 M_{\odot}$ are shot at hundreds of~$\kms$-- a
huge bundle that may be more accurately thought of as a flying wind
``bullet''.  An actual wind should form a bow shock resulting in a
plume of material spreading laterally, but instead in SPH this
speeding bullet is slowed down viscously through the interactions with
neighboring particles.  The number of interacting neighbors over the
same physical distance is lower at less resolution, and a wind
particle will take longer to slow down; this is an unfortunate but
unavoidable consequence of SPH.  Hence lower resolution simulations
inhibit recycling as evidenced by the average number of times an
individual wind particle is launched: 3.0, 2.5, and 2.1 times for 4.7,
38, and 300$\times10^6 M_{\odot}$ SPH particle resolution (to $z=0$).
This lack of resolution convergence is quantifiable by the median
recycling timescale of wind particles in different resolution
simulations as shown in the upper right panel of Figure
\ref{fig:recycle}, and is discussed in the following subsection.
Increased resolution suggests that wind recycling should move
$\Omega_{\rm wind}$ to an even higher value, probably exceeding
$1/2\Omega_b$.  Given that galaxies below a certain mass cannot form
owing to the presence of an ionizing background, there will be a limit
to how high $\Omega_{\rm wind}$ can be; in the future we hope to run
simulations that can achieve such resolutions in a cosmologically
representative volume.  For now, the wind recycling predictions should
be considered as illustrative rather than quantitative.

\subsection{Wind Recycling Timescales} \label{sec: galrecycle}

\begin{figure*}
\includegraphics[scale=0.80]{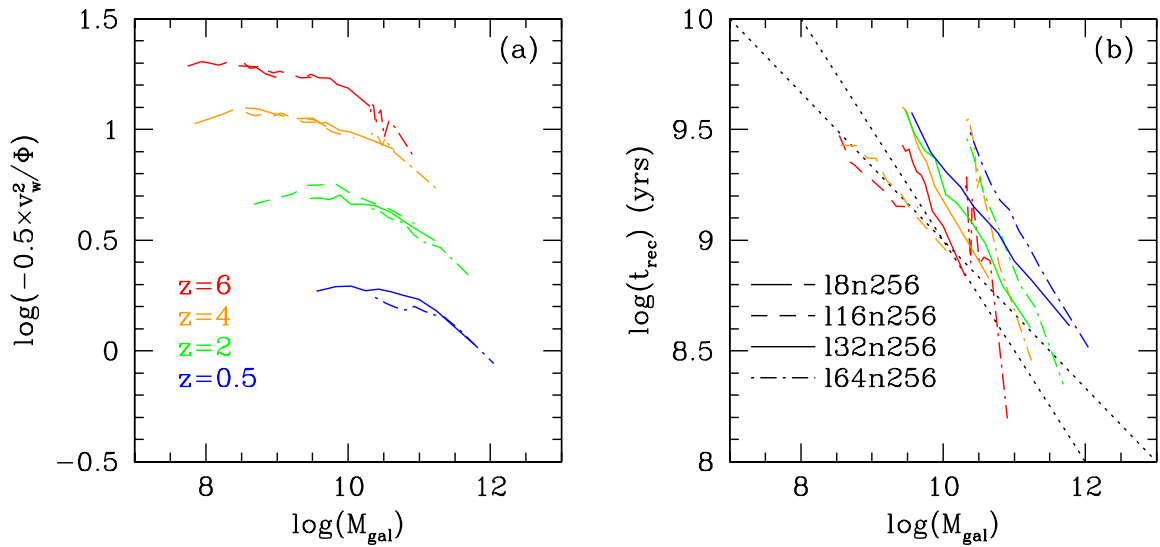}
\caption[]{The virial ratio ($E_{\rm rat}$, the ratio of wind kinetic
energy to potential well energy) are displayed as a function of galaxy
mass (derived by SKID) at 4 launch redshifts in Panel (a) for our
$\sigma$-derived wind simulations at various box sizes.  The shallow
turnover at high-mass is mostly due to imposed wind speed limits.
Panel (b) shows that the recycling timescale (the median time for a
wind particle to be relaunched in a wind) is a much stronger function
of $M_{\rm gal}$.  After exploring alternatives for the $M_{\rm
gal}-t_{\rm rec}$ anti-correlation, we conclude winds from more
massive galaxies encounter a denser environment slowing the wind
particle faster and allowing it to be re-accreted faster.  The two
dotted lines show the galaxy mass dependence of $t_{\rm rec}$ is
dominated by slowing due to environment, $M_{\rm gal}^{-1/3}$, or the
gravitational free-fall time, $M_{\rm gal}^{-1/2}$; wind particles
spend most of their time falling back into a galaxy.  The poor
resolution convergence in $t_{\rm rec}$ indicates lower resolution
simulations overpredict the $t_{\rm rec}$ and underpredict the total
amount of wind recycling.  To see wind recycling in action in our
simulations, see the movies and an explanation at
http://luca.as.arizona.edu/\~{}oppen/IGM/recycling.html.}
\label{fig:recycle}
\end{figure*}

We now examine wind recycling as a function of galaxy mass.  Specifically,
we want to know what galaxies are able to inject their metals into the
IGM.  The common thought is that low-mass galaxies lose their metals
more easily because winds can escape from these galaxies' shallower
potential wells~\citep[e.g.][]{dek86}.  As \citet{dek03} showed, if the
outflow energy couples efficiently to ambient halo gas, this can yield
the mass-metallicity relation as observed, as well as other properties
of dwarf galaxies.  However, \citet{fin08} showed that a model in
which galaxies expel material at constant velocity (i.e. the ``constant
wind" model of OD06) does not reproduce the observed mass-metallicity
relation, primarily because outflows do not in fact couple their energy
efficiently to ambient gas.  Instead, the observed mass-metallicity
relation~\citep[e.g.][]{erb06,tre04} are better reproduced in the
momentum-driven wind scenario.  In this case, since $\vw\propto
\sqrt{-\Phi_{\rm gal}}$ (roughly), all outflows have an approximately
equal probability of escaping their parent haloes independent of mass,
when considering only gravitational interactions with an isolated halo.

The left panel of Figure~\ref{fig:recycle} shows the virial ratio,
$E_{\rm rat}\equiv 0.5\vw^2/\Phi$, as a function of galaxy baryonic
mass, for winds launched at four redshifts.  At any given redshift,
the ratio is roughly constant, but rises slightly at low masses and
then declines to higher masses.  At low masses, the trend arises from
the mild metallicity dependence of $\vw$ together with the
mass-metallicity relation.  Moving to higher masses, since the
potential $\Phi$ includes the galaxy potential and the environmental
potential of the group/cluster in which it lies, and since massive
galaxies live in denser environments, $\Phi$ is greater.  At the
highest masses, our imposed speed limit reflecting the assumption that
wind energy cannot exceed twice the supernova energy reduces wind
velocities.  Note that by definition, $\Phi$-derived winds will have a
constant $E_{\rm rat}$ for a given launch redshift (modulo wind speed
limits); hence the curvature towards high masses is predominantly a
reflection of $\sigma$-derived winds.

In practice, outflows are not only confined by the gravitational potential
of the host galaxy; cosmic infall and hydrodynamic effects can also be
quite important.  For instance, \citet{fer05} used a simulation of a
typical LBG at $z=3$ ($M_{\rm dyn}=2\times10^{11} M_{\odot}$) to show that
metals never reach the IGM and instead are confined to the surrounding hot
halo gas by infalling gas that creates a shock interface with the outflow.

One way to quantify such additional effects is to consider a new
timescale called the recycling timescale, $t_{\rm rec}$, which is
defined as the median time for a wind particle to be re-ejected again
as a wind particle, or else fully converted into a star.  At all
launch redshifts explored, for all $M_{\rm gal}$, more than half of
wind particles recycle meaning that $t_{\rm rec}$ can be measured.
The right panel of Figure~\ref{fig:recycle} shows $t_{\rm rec}-M_{\rm
gal}$ relation, demonstrating a strong anti-correlation of $t_{\rm
rec}$ with galaxy mass.  The simple fact that winds recycle on short
timescales even though they are launched at velocities well above the
escape velocity is an indication that slowing by ambient gas and
surrounding potentials is significant\footnote{For new movies showing
wind recycling in action in our simulations, please visit
http://luca.as.arizona.edu/\~{}oppen/IGM/recycling.html}.

If the galaxy's gravity is dominant in confining its outflow, one
would expect that the time spent away from the galaxy would be
approximated by twice the free-fall timescale, $t_{\rm ff}$ (one
$t_{\rm ff}$ outward and another one back in, or simply the orbital
timescale for an orbit of eccentricity 1).  $t_{\rm ff}$ scales as
$R_{\rm turn}^{3/2}M_{\rm dyn}^{-1/2}$, where $R_{\rm turn}$ is the
turn-around distance of the wind particle, and $M_{\rm dyn}$ is the
dynamical mass of the galaxy treated as a point source for the sake of
simplicity.  Following Newtonian dynamics, $R_{\rm turn}\propto M_{\rm
dyn}/\vw^2$, and the $\vw\propto M_{\rm gal}^{1/3}$ relation for
momentum-driven winds, $t_{\rm ff}$ remains invariant as a function of
galaxy mass while $R_{\rm turn}\propto M^{1/3}$.  Momentum-driven
winds should reach a maximum distance from their parent galaxy that is
a constant multiple of the virial radius since $\vw$ scales with the
virial energy of an isolated halo.  The strong trend of $t_{\rm rec}$
with $M_{\rm gal}$ indicates that larger-scale potentials and
hydrodynamic effects are dominant.

Is the particle spending more time in the IGM/galactic halo or in the
galactic ISM before being relaunched?  The answer is that the vast
majority of the time between recycling is spent outside the star
forming regions of a galaxy.  This is not surprising, because in our
simulations the star formation timescale is tied completely to the
accretion timescale; once a particle is in a galaxy it gets converted
to a star or blown out in a wind much more quickly than it was
accreted~\citep{fin08}.  Therefore, we subdivide $t_{\rm rec}$ into two
timescales, one leaving the galaxy to reach $R_{\rm turn}$ while being
slowed by hydrodynamical forces, $t_{\rm out}$ and the other returning to
the galaxy in approximately one $t_{\rm ff}$.  For simplicity, let's say
the hydrodynamic forces slow the wind particle down at a constant
rate, in which case $t_{\rm out}=R_{\rm turn}/(\vw/2)$.  The total recycling
timescale can be approximated as
\begin{eqnarray}\label{eqn:trec}
\frac{t_{\rm rec}}{1 \rm{Gyr}} = 1.96 \left(\frac{R_{\rm turn}}{100 \rm{kpc}}\right) \left(\frac{100 \kms}{\vw}\right) 
\nonumber\\+~2.09 \left(\frac{R_{\rm turn}}{100 \rm{kpc}}\right)^{3/2} \left(\frac{10^{10} M_{\odot}}{M_{\rm gal}}\right)^{1/2}
\end{eqnarray}
if we assume that $M_{\rm dyn}=\Omega_{M}/\Omega_{b}\times M_{\rm gal}$ and
$R_{\rm turn}$ is in physical kpc.  We can now solve for $R_{\rm turn}$ by
taking the $t_{\rm rec}$ from Figure~\ref{fig:recycle} and the
average $\vw$ from Figure~\ref{fig:vwind}.  $R_{\rm turn}$ is just an
upper limit for how far wind particles with median $t_{\rm rec}$ can
extend from a parent galaxy, because in reality a particle does not
likely go out and then immediately fall back into a galaxy in two
steps; wind particles often spend time orbiting around their parent
galaxies as well.

In principle, $R_{\rm turn}$ can be tracked directly in simulations,
however our current suite of runs did not output position information
of wind particles owing to the large storage requirements.  We have
checked for isolated cases that our simple formula (eq.~\ref{eqn:trec})
yields roughly correct results.

\begin{figure*}
\includegraphics[scale=0.80]{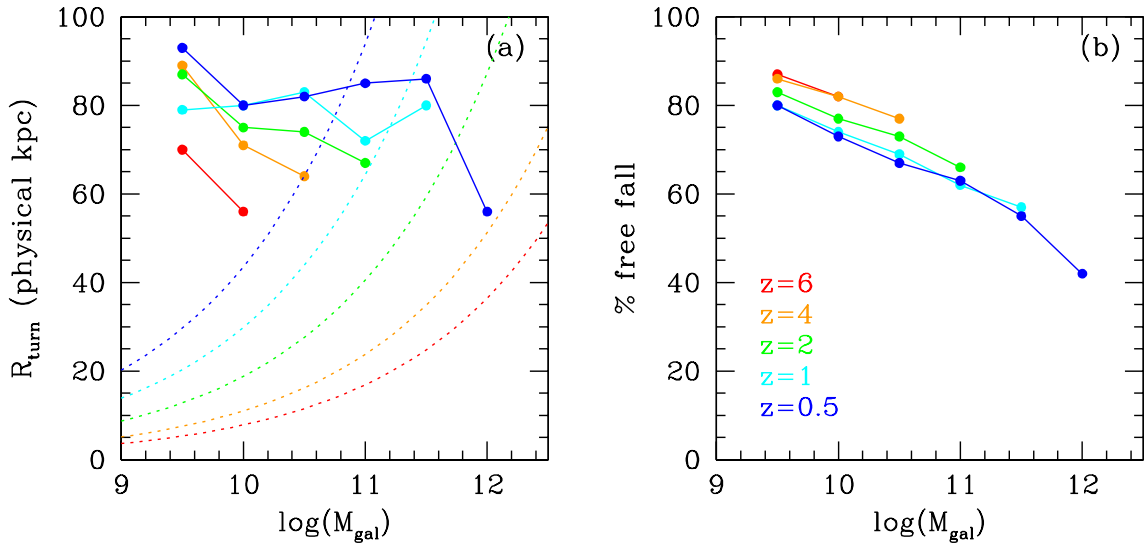}
\caption[]{Panel (a) shows $R_{\rm turn}$, the maximum radial extent a
median wind particle may extend from its parent galaxy calculated from
equation~\ref{eqn:trec} in the l32n256 simulation.  $R_{\rm turn}$
generally is larger at smaller $M_{\rm gal}$; this trend coupled with the
fact that $t_{\rm rec}$ is longer indicates that smaller galaxies can more
easily enrich the IGM.  Dotted lines corresponding to a NFW halo
radius, $r_{\rm 200}$ \citet{nav97}, show that winds from $L^*$ galaxies
do not escape their parent haloes at $z<1$.  Panel (b) shows that
winds from small galaxies spend more time on their journey returning
to a galaxy than being blown out.  }  \label{fig:rturn}
\end{figure*}

We plot $R_{\rm turn}$ in the left of Figure~\ref{fig:rturn}, focusing
first on the trend over 2.5 decades of galaxy masses at $z=0.5$, which
we consider here the local Universe since trends do not evolve much
until $z=0$.  $R_{\rm turn}$ stays nearly constant from the size scale of
dwarf galaxies to galaxies above $L^*$ while $t_{\rm rec}$ declines.
Dotted lines correspond to the radius, $r_{\rm 200}$, of a NFW halo
\citep{nav97}.  The point at which $R_{\rm turn}$ and $r_{\rm 200}$ intersect
is the approximate transition mass below which winds reach the IGM and
above which winds are confined to their haloes.  Metals rarely enter
the IGM from galaxies above $10^{11} M_{\odot}$ in the local Universe.
Several significant conclusions can be drawn from this behavior at
low-$z$.

First let us consider our wind model and the Milky Way, which has
$M_{\rm gal}\sim 10^{11} M_{\odot}$.  Our wind model produces an
outflow from a Milky Way-type galaxy, with initial speeds of a few
hundred~$\kms$.  One might claim that this immediately invalidates our
model, since the Milky Way is not observed to have an outflow.  Yet
quantitatively, our model predicts that a Milky Way-sized galaxy
should have a recycling time of less than 1~Gyr, and the typical wind
particle will not venture beyond 85 kpc (if we assume not much has
changed between $z=0.5\rightarrow0$).  In other words, in our model,
the Milky Way is not driving a classical outflow as seen from local
starbursts, but rather is continually sending material up into the
halo and having it rain back down on a timescale of 1~Gyr.  We call
this a {\it halo fountain}, in analogy with a galactic fountain that
operates on smaller scales.  Indeed, we speculate that there may not
be a fundamental difference between galactic fountains and halo
fountains, but rather gas is being thrown out of the disk at a range
of velocities; however, our simulations lack the resolution to address
this issue directly.

How might such a halo fountain be observed?  One possibility is that
it has already been seen, as high velocity clouds (HVC's).
\citet{wak97} predict that given the observed rate of material going
into HVC's of about 5 $M_{\odot} yr^{-1}$ then it should take $\sim1$
Gyr for all the gas in the ISM to cycle through this halo fountain.
Hence the Galaxy may have an active halo fountain recycling its
material on a timescale much less than a Hubble time, despite not
resembling anything like a starburst galaxy.  The difficulty of
observation of feedback within our own Galaxy may just indicate how
invisible yet ubiquitous galactic-scale winds are.

Secondly, with $R_{\rm turn}$ nearly constant, $t_{\rm rec}$ should be
proportional to $M_{\rm gal}^{-1/3}$ if dominated by $t_{\rm out}$ and
$M_{\rm gal}^{-1/2}$ if $t_{\rm ff}$ is larger.  The black dotted line in the
right of Figure~\ref{fig:recycle} corresponding to the latter case
(the steeper of the two) appears to more closely match the general
trends indicating that $t_{\rm ff}$ dominates; wind particles spend a
majority of $t_{\rm rec}$ falling into galaxies (Panel (b) in
Figure~\ref{fig:rturn}).  Our calculations above agree $t_{\rm ff}$ grows
larger than $t_{\rm out}$ for smaller galaxies, but the two timescales are
similar for extremely massive galaxies, $M_{\rm gal}\sim10^{12}
M_{\odot}$.

\begin{figure}
\includegraphics[scale=0.90]{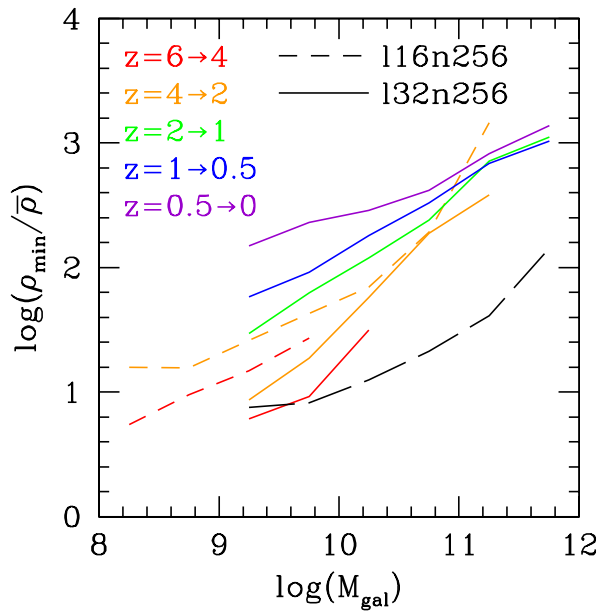}
\caption[]{The minimum density a wind particle achieves between
recyclings is shown to be a strong function of galaxy mass at all
redshifts (colored lines).  Lower mass galaxies have longer recycling
times, because winds are launched into less dense environments where
shocks take longer to slow and turn around winds.  The environmental
density within sphere of 1 comoving Mpc as a function of $M_{\rm gal}$ at
$z=0.5$ (black dashed line from l32n256 simulation) shows the same
trend as $\rho_{\rm min}/\bar{\rho}$ indicating that environment is the
primary factor in how far winds reach into the IGM.  The NFW halo
profile alone cannot explain this trend as halo profiles drop much
more sharply and the environmental density dominates for all but the
most massive galaxies at low-$z$.}
\label{fig:minrho}
\end{figure}

Thirdly, smaller galaxies live in less dense environments where
hydrodynamic slowing takes longer.  To demonstrate this we look at
another parameter, the minimum overdensity reached by wind particles,
which should approximately correspond to the density at $R_{\rm
turn}$.  For the subset of wind particles we track the minimum density
achieved before recycling, $\rho_{\rm min}$.  Figure \ref{fig:minrho}
(solid lines) shows the median $\rho_{\rm min}$ as a function of the
baryonic mass of the originating galaxy, $M_{\rm gal}$, from the
l32n256 run.  At all redshifts, smaller galaxies push their winds to
lower densities, consistent with them having longer recycling times.
Their less dense environments slow winds over a longer $t_{\rm out}$,
and allow them to reach a similar $R_{\rm turn}$ as more massive
galaxies despite lower $\vw$.  We also plot a long-dashed line that
show the average density within 1~comoving Mpc sphere around the
$z=0.5$ galaxies.  These show a similar trend, indicating that
environmental dependence is the primary factor in how far winds reach
into the IGM and how long they remain there.  The fact that $\rho_{\rm
min}$ is much higher than the density within 1 Mpc is an indication
that these winds are traveling much less than that distance as our
calculations above indicate ($\sim 80$ kpc comoving at $z=0.5$).

Although a possible explanation for the trend in $\rho_{\rm min}$ is
larger haloes have higher densities at the same distance, this cannot
account for the dependence.  The density of NFW haloes decline nearly
as $1/r$ beyond $r_{200}$ resulting in a $\rho_{\rm min}\propto
M_{gal}$ dependence much steeper than plotted in
Figure~\ref{fig:minrho} below $z=2$; for all but the most massive
galaxies the density contribution of the parent halo is much smaller
at $\rho_{\rm min}$.  The flattening of this relation toward $z=0$ at
$M_{gal}<10^{11} M_{\odot}$ means these galaxies more likely live in
denser environments.  For more massive galaxies, the parent halo
itself is more responsible for the hydrodynamical slowing.

Finally, small galaxies can enrich a similar volume as large galaxies
leading to major implications as to which galaxies enrich the IGM.
Metals from small galaxies have the advantage of staying in the IGM
longer, however the disadvantage of their winds being less metal
enriched (e.g. Panel (b) of Figure~\ref{fig:mfunc}).  We plan to look
at the origin of IGM metal absorbers in a future paper, exploring what
galaxies enrich the IGM, how long these metals have been in the IGM,
and what distance metals travel from their parent galaxy.

As a side note, we admit that the global SFRD is overestimated at late
times ($z<0.5$); the l32n256 run showing nearly 4$\times$ as much SF
at $z=0$ relative to observations (Figure~\ref{fig:global}).  The
primary reason for this discrepancy is the lack of a quenching
mechanism in the largest galaxies; galaxies above $10^{11} M_{\odot}$
account for 54\% of star formation at $z=0$, and their continued late
growth creates far too many galaxies above this mass as compared to
observations \citep{per08}.  The influence of these over-massive
galaxies on the IGM however appears to be remote due to their
inability to inject metals beyond their haloes and their short
recycling timescales. 

Turning to the evolution of recycling, $t_{\rm rec}$ grows moderately
longer at lower redshift for a given $M_{\rm gal}$.  For a $10^{10}
M_{\odot}$ galaxy, $t_{\rm rec}$ is 1.1, 1.5, 1.6, and 2.0 Gyr for
launch redshifts 6, 4, 2, and 0.5 respectively in the 32 $\hmpc$ box.
This trend is in place despite $E_{\rm rat}$ declining by 1 decade
from $z=6\rightarrow0.5$ (see Figure \ref{fig:recycle}(a)); wind
kinetic energy is declining relative to the potential at their launch
location.  Again, the overriding variable is the slowing of the wind
by the environment; the average physical density declines by a factor
of 100 in this interval, making it easier for winds to travel further
despite a factor of $10\times$ less energy input into the winds.
Galaxies of the same mass are more likely to live at higher
overdensities at lower redshift as structural growth makes more groups
and clusters, but the declining physical densities and slower Hubble
expansion of the Universe toward low redshift outweigh this.

While the $t_{\rm rec}$ grows at low redshift, the amount of the
Universe it enriches sharply declines.  The calculations of $R_{\rm
turn}$ for equation~\ref{eqn:trec} shows a moderately increasing
physical distance from $z=6\rightarrow0.5$ with values of 56, 71, 75,
and 80 kpc for the four respective launch redshifts when considering
the same $10^{10} M_{\odot}$ galaxy.  However, the comoving volume
such a galaxy is enriching is {\it 35$\times$ greater} at $z=6$ than
at $z=0.5$.  Momentum-driven winds can more easily expel metals into
the IGM at high redshift, whereas the overdense environments combined
with weaker wind velocities and the overall larger scale of the
Universe leave metals perpetually recycling in halo fountains in
groups and clusters at low redshift.  These are the {\it primary}
reasons metals migrate from the IGM to galaxies as the Universe
evolves.

Our results compare favorably to those of \citet{ber07}, who follow
energy-driven winds ($\eta\propto\sigma^{-2}$ and $\vw\propto\sigma$)
in their semi-analytical implementation using a two-phase process for
the dynamical wind evolution (see also \citet{del07}).  At $z=0$ they
also find that it is the smaller galaxies ($10^{10.5} M_{\odot}<M_{\rm
dyn}<10^{11.5} M_{\odot}$, i.e. $\sim 1/10 M^{*}$) that most
efficiently inject their metals into the IGM ($\sim40$\% of the time),
while metals from more massive haloes than $10^{12} M_{\odot}$ rarely
escape the halo and are recycled (see their Figure 9).  This is
despite their $\vw$-dependence (same as ours but without the scatter)
exceeding the escape velocity.  The redshift dependence is harder to
compare to our results due to differing definitions of escaped
fractions into the IGM, but they do also see smaller haloes, the
progenitors to massive galaxies today, as the most efficient enrichers
of the IGM up to $z=4.5$.

Lastly, $t_{\rm rec}$ and $\rho_{\rm min}$ show similar resolution
convergence issues as $\Omega_{\rm wind}$, for similar reasons.  Lines
at the same redshift do not overlap in either of these plots,
indicating the $t_{\rm rec}$ is longer for the same $M_{\rm gal}$ at lower
resolution.  As discussed before, coarser resolution simulations slows
down particles over a longer physical distance allowing wind particles
to travel artificially too far at low resolution; this is why there is
less recycling at lower resolution.  For well-resolved galaxies, the
lines of different resolution appear to nearly converge.  The
hydrodynamical treatment of wind particles after they are launched is
not nearly as resolution-converged as the feedback properties on per
galaxy basis (see Figure \ref{fig:mfunc}).  Recycling times are
overestimated when under-resolved.  Hence recycling times may actually
be shorter than we've shown.

In summary, we have shown that wind recycling is an important
phenomenon for understanding the evolution of galaxies and the IGM.
Winds from galaxies are typically expelled and re-accreted on
timescales short compared to a Hubble time.  This occurs despite the
fact that winds are typically launched with plenty of kinetic energy
to escape its halo.  Such gravitational arguments are therefore not
very relevant to understand how material cycles through galaxies and
the IGM; instead, infall and shocks generated by outflows are more
important in setting the wind distribution length- and time-scales.
The recycling time is therefore strongly anti-correlated with galaxy
mass, owing to the fact that more massive galaxies live in denser
environments.  These denser environments make it harder for winds to
travel beyond their parent halo at late times causing the migration of
metals from the IGM to galaxies from $z=2\rightarrow 0$.  At the
present epoch, small galaxies can still expel material, but larger
galaxies are more aptly described as having halo fountains, in which
material is constantly kicked up into the halo before raining down.

\section{Summary} \label{sec: summary}

We introduce a new version of \gad, with improvements designed to
explore mass, metal, and energy feedback from galactic outflows across
all cosmic epochs.  We add two major modules designed to make the code
better suited to explore the low-$z$ Universe: (1) A sophisticated
enrichment model tracking four elements individually from Type II SNe,
Type Ia SNe, and AGB stars; and (2) an on-the-fly galaxy finder used
to derive momentum-driven wind parameters based on a galaxy's mass.

We first run test simulations to explore global energy and enrichment
properties with and without AGB and Type Ia feedback, and with our old
and new wind implementations.  Focusing on our new (galaxy
mass-derived) winds including all sources of feedback, we then run
several 34 million-particle simulations to explore feedback over the
history of the Universe and over a large dynamic range in galaxy mass.
We also track a representative subset of wind particles to study in
detail how mass, metals, and energy are distributed by outflows.

Our new chemical enrichment model enables us to investigate the global
production and distribution of key individual metal species.
Globally, metal production of all four species tracked (C, O, Si, Fe)
is dominated by Type~II SNe at all redshifts.  Type~Ia SNe add
significantly to the iron content of hot, intracluster gas, especially
at $z<1$.  AGB stars add moderately to the IGM carbon abundance by
$z=0$, and provide fresh (enriched) gas for recycling into stars which
increases global star formation at later epochs.  Carbon yields from
AGB stars cannot be ignored even at high-$z$, because carbon AGB stars
enrich on timescales much less than a local Hubble time (i.e. 200
Myr-1 Gyr).  Due to the complex interplay between instantaneous and
delayed recycling from various forms, metallicity patterns in the IGM
and ICM cannot be straightforwardly used to infer the enrichment
patterns in the host galaxies responsible for polluting intergalactic
gas.

We study enrichment patterns subdivided by baryonic phase.  The total
metal mass density is split roughly equally between galaxies and the
IGM from $z=6\rightarrow 2$.  Shocked intergalactic gas (WHIM and ICM)
contains a fairly small portion of the global metal mass at all
epochs.  At $z<2$, metals tend to migrate from the IGM into galaxies,
so that by $z=0$ about two-thirds of the metals are in galaxies
(i.e. stars and cold gas).  The combination of increased carbon and
gas recycling from AGB feedback results in the IGM being significantly
more carbon-enriched, which helps reproduce the relatively high
observed mass density in \CIV\ absorption systems at $z\approx 0$.

Our new galaxy mass-based wind model implementation provides a more
faithful representation of the momentum-driven wind model of
\citet{mur05}, and yields outflows that are in better broad agreement
with observed outflows.  In particular, our implementation results in
faster winds at high-$z$ and slower wind speeds at low-$z$ compared to
our old local potential-derived winds.  The fast early winds are able
to enrich the IGM at early times as observed, while the slow late
winds mean that most galaxies today are not driving material into the
IGM at all.  Qualitatively, this better agrees with observations
indicating that most galaxies at $z\sim 2-3$ drive powerful
winds~\citep{erb06}, while today galaxies are rarely seen to have
strong outflows.  

We examine bulk properties of outflows as a function of galaxy mass.
We find that mass, metallicity, and energy feedback as a function of
galaxy baryonic mass roughly follow the trends predicted by
momentum-driven winds: $\dot{M}_{\rm wind}\propto M_{\rm gal}^{2/3}
H^{-1/3}(z)$ and $\dot{E}_{\rm wind}\propto M_{\rm gal}^{4/3}
H^{1/3}(z)$.  The metal outflow rate is $\dot{M}_{\rm wind} Z_{\rm
gal}$, where $Z_{\rm gal}\propto M_{\rm gal}^{1/3}$ with a
proportionality constant that increases with time as given by
mass-metallicity evolution~\citep[cf.][]{fin08}.  The stellar mass of
the typical (median) galaxy most responsible for each particular form
of feedback increases by a factor of $\sim30$ between
$z=6\rightarrow2$, but only $\sim\times 2-3$ from $z=2\rightarrow0$.
Each form of feedback traces a different mass scale: mass feedback
$\sim\frac{1}{6} M^{*}$, energy feedback $\sim M^{*}$, and metallicity
feedback $\sim\frac{1}{2} M^{*}$.  Measuring these characteristic
masses~\citep[e.g.][]{bou07} offers the possibility to test whether
momentum-driven wind scalings are followed globally.

The wind energy relative to the supernova energy scales roughly as
$M_{\rm gal}^{1/3}$, and is within a factor of a few of unity at all
epochs and galaxy masses.  Given expected radiative losses from SN
heat input into the ISM, it seems that SNe will have a difficult time
providing enough energy to pollute the Universe as observed.  An
alternate source of energy would ease this tension, such as photons
from young stars whose total energy can exceed the supernova energy by
several orders of magnitude.  This adds to the circumstantial evidence
supporting the idea that galactic outflows may be driven in large part
by radiation pressure, as postulated in the momentum-driven wind
model.

We find that {\it wind recycling}, material ejected as outflows and
then re-accreted and ejected again, turns out to be a remarkably
common occurrence with significant dynamical repercussions.  By
following individual wind particles in our simulations down to $z=0$,
we find that multiple recyclings are the norm, and that the typical
wind particle has been ejected three to four times.  In other words,
outflow material being reaccreted onto a galaxy dominates over outflow
material launched into the IGM forever.  Since in our wind model all
outflows are launched at speeds well above the escape velocity of
galaxies, this indicates that outflows are mainly slowed through
hydrodynamic interactions, allowing them to rejoin the hierarchical
accretion flow into galaxies.  Approximately 20\% of the baryons
participate in an outflow, but owing to multiple launchings the
aggregate mass of baryons ejected exceeds half of the total baryonic
mass.  The two key corollaries of wind recycling as seen in our
simulations are: (1) Material driven in an outflow cannot be assumed
to remain in the IGM forever, and (2) Gravitational energetic
considerations are generally not relevant for determining how far
outflow propagate into the IGM.

We examine wind recycling as a function of galaxy mass.  The recycling
time scales roughly as $t_{\rm rec}\propto M_{\rm gal}^{-1/2}$, as
expected if environmental effects are dominating the retardation of
outflows.  Larger galaxies have shorter recycling times despite
launching winds at larger speeds, because they live in denser
environments; the minimum overdensity achieved by winds scales with
galaxy mass.  As expected from the ubiquity of multiple recyclings,
the recycling time is generally fairly short, roughly $10^{9\pm
0.5}$~years, increasing only mildly at lower redshift.  Our analysis
suggest that winds generally return to the galaxy from which they were
launched (or its descendant), though a full merger tree construction
is required to confirm this.

It is possible to estimate how far outflows travel from their host
galaxies ($R_{\rm turn}$), as a function of galaxy mass and redshift.
Remarkably, $R_{\rm turn}=80\pm 20$~{\it physical} kpc at all
redshifts and masses.  There are weak trends for higher $R_{\rm turn}$
at smaller masses and lower redshifts; both are consistent with
ambient density being a key determinant for how far winds travel.  The
constant physical distance means that outflows at early epochs are
able to enrich a significant fraction of the Universe, while outflows
at later epochs are more confined around galaxies.  This is the reason
metals migrate from the IGM to galaxies between $z=2\rightarrow 0$.

Comparing $R_{\rm turn}$ to halo radii, we see that at $z>2$, typical
$L^*$-sized galaxies have outflows that escape their host halos into
the IGM, while at $z<1$ outflows are generally confined within
galactic halos.  This gives rise to the concept of {\it halo
fountains}, where low-$z$ galaxies are constantly kicking gas out of
their ISM into the halo but no further, and this material rains back
down onto the ISM on timescales of order 1~Gyr or less.  If correct,
then even galaxies not canonically identified as having outflows (such
as the Milky Way) may in fact be moving a significant amount of
material around its halo.  This halo fountain gas would be quite
difficult to detect, as it is likely to be tenuous and multi-phase; we
broadly speculate it might be responsible for high-velocity clouds or
halo MgII absorbers~\citep[e.g.][]{kac08}.  We leave a more thorough
investigation of the observational consequences of halo fountains for
the future.

As a final caveat, it should be pointed out that detailed properties
of how outflows propagate out of galaxies are not as well-converged
with numerical resolution as we would like.  This may be due to our
particular way of implementing outflows in a Monte Carlo fashion
combined with difficulties of SPH in handling individual outflowing
particles.  Here we have focused on qualitative trends that appear to
be robust within our limited exploration of resolution convergence,
without making overly detailed quantitative predictions.  In the
future we plan to investigate how to implement outflows in a more
robust way within the framework of cosmological hydrodynamic
simulations.

Wind recycling and halo fountains provide two new twists on the idea
of galactic outflows.  If our models are correct, then there is a
continuum of outflow properties from galactic fountains that barely
kick gas out of the disk, to halo fountains that cycle material
through gaseous halos of galaxies, to large-scale outflows that are
e.g. responsible for enriching the IGM.  In short, outflows may be
considerably more ubiquitous and complicated than previously thought,
and hence understanding their effects on galaxies as a function of
mass, environment, and epoch will be even more critical for developing
a comprehensive model for how galaxies form and evolve.

\section*{Acknowledgments}  \label{sec: ack}

We thank J. Bieging, N. Bouch\'e, M. Fardal, K. Finlator, N. Katz,
and R. Somerville for useful conversations and inspiring new avenues of
exploration.  We thank V. Springel for contributing his FOF finder, which
we later modified for use in our simulations. The simulations were run on
the Xeon Linux Supercluster at the National Center for Supercomputing
Applications, the University of Arizona's 512-processor SGI Altix
system housed at the Center for Computing \& Information Technology,
and our 16-processor Opteron systems.  The authors thank H.-W. Rix and
the Max-Planck Instit\"ut f\"ur Astronomie for their gracious hospitality
while some of this work was being done.  Support for this work, part of
the Spitzer Space Telescope Theoretical Research Program, was provided
by NASA through a contract issued by the Jet Propulsion Laboratory,
California Institute of Technology under a contract with NASA.  Support
for this work was also provided by NASA through grant number HST-AR-10946
from the SPACE TELESCOPE SCIENCE INSTITUTE, which is operated by AURA,
Inc. under NASA contract NAS5-26555.

\label{lastpage}


\begin{thebibliography}{}

\bibitem[\protect\citeauthoryear{Anders \& Grevesse} {1989}]{and89} Anders, E. \& Grevesse, N. 1989, Geochim. Cosmochim. Acta, 53, 197

\bibitem[\protect\citeauthoryear{Aguirre et al.} {2001}]{agu01} Aguirre, A., Hernquist, L., Schaye, J., Weinberg, D. H., Katz, N., Gardner, J. 2001, ApJ, 560, 599

\bibitem[\protect\citeauthoryear{Balogh et al.} {2001}]{bal01} Balogh, M. L., Pearce, F. R., Bower, R. G., Kay, S. T. 2001, MNRAS, 326, 1228

\bibitem[\protect\citeauthoryear{Bertone et al.} {2007}]{ber07} Bertone, S., De Lucia, G., \& Thomas, P. A. 2007, MNRAS, 379, 1143

\bibitem[\protect\citeauthoryear{Bell et al.} {2003}]{bel03} Bell, E. F., McIntosh, D. H., Katz, N., \& Weinberg, M. D. 2003, ApJS, 149, 289

\bibitem[\protect\citeauthoryear{Bokensberg, Sargent, \& Rauch} {2003}]{bok03} Boksenberg, A., Sargent, W. L. W., \& Rauch, M. 2003, ASP Conference Proceedings, Vol. 297, 447, eds. Edited E. Perez, R.M.G. Delgado, \& G. Tenorio-Tagle (BSR03)

\bibitem[\protect\citeauthoryear{Bouch\'e et al.} {2007}]{bou07} Bouch\'e, N., Lehnert, M. D., Aguirre, A., P\'eroux, C., \& Bergeron, J. 2007, MNRAS, 378, 525 (B07)

\bibitem[\protect\citeauthoryear{Bouwens et al.} {2007}]{bouw07} Bouwens, R. J., Illingworth, G. D., Franx, M., \& Ford, H. 2007, ApJ, 670, 928

\bibitem[\protect\citeauthoryear{Bressan et al.} {1993}]{bre93} Bressan A., Fagotto F., Bertelli G., \& Chiosi C. 1993, A\&AS, 100, 647

\bibitem[\protect\citeauthoryear{Bruzual \& Charlot} {2003}]{bru03} Bruzual, G. \& Charlot, S. 2003, MNRAS, 344, 1000

\bibitem[\protect\citeauthoryear{Caputi et al.} {2006}]{cap06} Caputi, K. I. et al. 2006, ApJ, 637, 727

\bibitem[\protect\citeauthoryear{Cattaneo et al.} {2007}]{cat07} Cattaneo, A., Blaizot, J., Weinberg, D. H., Colombi, S., Dav\'e, R., Devriendt, J., Guiderdoni, B., Katz, N., Keres, D. 2007, MNRAS, 377, 63.

\bibitem[\protect\citeauthoryear{Cen \& Ostriker} {1999}]{cen99} Cen, R. \& Ostriker, J. P., 1999, ApJ, 514, 1

\bibitem[\protect\citeauthoryear{Cen et al.} {2001}]{cen01} Cen, R., Tripp, T. M., Ostriker, J. P., Jenkins, E. B. 2001, ApJL, 559, L5

\bibitem[\protect\citeauthoryear{Cen \& Ostriker} {2006}]{cen06a} Cen, R. \& Ostriker, J. P. 2006, ApJ, 650, 560

\bibitem[\protect\citeauthoryear{Cen \& Fang} {2006}]{cen06b} Cen, R. \& Fang, T. 2006, ApJ, 650, 573

\bibitem[\protect\citeauthoryear{Chabrier} {2003}]{cha03} Chabrier G., 2003, PASP, 115, 763

\bibitem[\protect\citeauthoryear{Chen et al.} {2003}]{che03} Chen, X., Weinberg, D. H., Katz, N., Dav\'e, R. 2003, ApJ, 594, 42

\bibitem[\protect\citeauthoryear{Cole et al.} {2001}]{col01} Cole, S. et al. 2001, MNRAS, 326, 255

\bibitem[\protect\citeauthoryear{Cowie \& Songaila} {1998}]{cow98} Cowie, L. L. \&  Songaila,A. 1998, Nature, 394, 44

\bibitem[\protect\citeauthoryear{Dalla Vecchia \& Schaye} {2008}]{dal08} Dalla Vecchia, C. \& Schaye, J. 2008, astro-ph/0801.2770


\bibitem[\protect\citeauthoryear{Dav\'e et al.} {1999}]{dav99} Dav\'e, R., Hernquist, L., Katz, N., \& Weinberg, D. H. 1999, ApJ, 511, 521

\bibitem[\protect\citeauthoryear{Dav\'e, Finlator \& Oppenheimer} {2006}]{dav06} Dav\'e, R., Finlator, K., \& Oppenheimer, B. D. 2006, MNRAS, 370, 273

\bibitem[\protect\citeauthoryear{Dav\'e \& Oppenheimer} {2007}]{dav07} Dav\'e, R. \& Oppenheimer, B. D. 2006, MNRAS, 374, 427 (DO07)

\bibitem[\protect\citeauthoryear{Dav\'e} {2008}]{dav08a} Dav\'e, R. 2008, MNRAS, accepted, arXiv:0710.0381

\bibitem[\protect\citeauthoryear{Dav\'e, Sivanandam \& Oppenheimer} {2008}]{dav08b} Dav\'e. R., Sivanandam, S., Oppenheimer, B. D. 2008, in preparation

\bibitem[\protect\citeauthoryear{De Lucia \& Blaizot} {2007}]{del07} De Lucia, G. \& Blaizot, J. 2007, MNRAS, 375, 2

\bibitem[\protect\citeauthoryear{Dekel \& Silk} {1986}]{dek86} Dekel, A. \& Silk, J. 1986, ApJ, 303, 39

\bibitem[\protect\citeauthoryear{Dekel \& Woo} {2003}]{dek03} Dekel, A., \& Woo, J.\ 2003, MNRAS, 344, 1131

\bibitem[\protect\citeauthoryear{Dekel \& Birnboim} {2006}]{dek06} Dekel, A. \& Birnboim, Y. 2006, MNRAS, 368, 2

\bibitem[\protect\citeauthoryear{Di Matteo, Springel \& Hernquist} {2005}]{dim05} Di Matteo, T., Springel, V., \& Hernquist, L. 2005, Nature, 433, 604

\bibitem[\protect\citeauthoryear{Erb et al.} {2006}]{erb06} Erb, D. K., Shapley, A. E., Pettini, M., Steidel, C. C., Reddy, N. A., \& Adelberger, K. L. 2006, ApJ, 644, 813 

\bibitem[\protect\citeauthoryear{Evrard et al.} {2008}]{evr08} Evrard, A. E., et al. 2008, ApJ, 672, 122

\bibitem[\protect\citeauthoryear{Fardal et al.} {2007}]{far07} Fardal, M. A., Katz, N., Weinberg, D. H., Dav\'e, R. 2007, MNRAS, 379, 985.

\bibitem[\protect\citeauthoryear{Ferrara et al.} {2005}]{fer05} Ferrara, A., Scannapieco, E., \& Bergeron, J. 2005, ApJ, 634, L37

\bibitem[\protect\citeauthoryear{Finlator et al.} {2006}]{fin06} Finlator, K., Dav\'e, R., Papovich, C., \& Hernquist, L. 2006, ApJ, 639, 672

\bibitem[\protect\citeauthoryear{Finlator et al.} {2007}]{fin07a} Finlator, K., Dav\'e, R., \& Oppenheimer, B. D. 2007, MNRAS, 376, 1861

\bibitem[\protect\citeauthoryear{Finlator \& Dav\'e} {2008}]{fin08} Finlator, K. \& Dav\'e, R., 2008, MNRAS, accepted 

\bibitem[\protect\citeauthoryear{Fontana et al.} {2006}]{fon06} Fontana, A. et al. 2006, A\&A, 459, 745

\bibitem[\protect\citeauthoryear{Frye et al.} {2003}]{fry03} Frye, B. L., Tripp, T. M., Bowen, D. B., Jenkins, E. B., \& Sembach, K. R. 2003, in ``The IGM/Galaxy Connection: The Distribution of Baryons at z=0'', ASSL Conference Proceedings Vol. 281, 231, eds. J.L. Rosenberg \& M.E. Putman

\bibitem[\protect\citeauthoryear{Fujita et al.} {2004}]{fuj04} Fujita, A., Mac Low, M.-M., Ferrara, A., Meiksin, A. 2004, ApJ, 613, 159

\bibitem[\protect\citeauthoryear{Fujita et al.} {2008}]{fuj08} Fujita, A., Martin, C. L., Mac Low, M.-M., New, K. C. B, \& Weaver, R. 2008, arXiv:0803.2892, submitted to ApJ

\bibitem[\protect\citeauthoryear{Fukugita \& Peebles} {2004}]{fuk04} Fukigita, M. \& Peebles, P. J. E. 2004, ApJ, 616, 643

\bibitem[\protect\citeauthoryear{Gavil\'an et al.} {2005}]{gav05} Gavil\'an M., Buell J. F., Moll\'a M. 2005, A\&A, 432, 861

\bibitem[\protect\citeauthoryear{Helsdon \& Ponman} {2000}]{hel00} Helsdon, S. F. \& Ponman, T. J. 2000, MNRAS, 315, 356

\bibitem[\protect\citeauthoryear{Herwig} {2004}]{her04} Herwig, F. 2004, ApJS, 155, 651

\bibitem[\protect\citeauthoryear{Hirschi et al.} {2005}]{hir05} Hirschi, R., Meynet, G., \& Maeder, A. 2005, A\&A, 433, 1013

\bibitem[\protect\citeauthoryear{Hopkins \& Beacom} {2006}]{hop06} Hopkins, A. M. \& Beacom, J. F. 2006, ApJ, 651, 142


\bibitem[\protect\citeauthoryear{Kacprzak et al.} {2008}]{kac08} Kacprzak, G. G., Churchill, C. W., Steidel, C. C., Murphy, M. T. 2008, AJ, 135, 922

\bibitem[\protect\citeauthoryear{Kennicutt} {1998}]{ken98} Kennicutt, R. C. 1998, ApJ, 498, 541

\bibitem[\protect\citeauthoryear{Kere\v{s} et al.} {2005}]{ker05} Kere\v{s}, D., Katz, N., Weinberg, D. H., \& Dav\'e, R. 2005, MNRAS, 363, 2

\bibitem[\protect\citeauthoryear{Kobayashi} {2004}]{kob04} Kobayashi, C. 2004, MNRAS, 347, 740

\bibitem[\protect\citeauthoryear{Kobayashi et al.} {2007}]{kob07} Kobayashi, C., Springel, V., \& White, S. D. M. 2007, MNRAS, 376, 1465

\bibitem[\protect\citeauthoryear{Lia et al.} {2002}]{lia02} Lia C., Portinari L., Carraro G., 2002, MNRAS, 330, 821

\bibitem[\protect\citeauthoryear{Lilly et al.} {1996}]{lil96} Lilly, S. J., Le Fevre, O., Hammer, F., Crampton, D. 1996, ApJL, 460, L1 

\bibitem[\protect\citeauthoryear{Limongi \& Chieffi} {2005}]{lim05} Limongi, M. \& Chieffi, A. 2005, ASP Conference Series, Vol. 342, 1604-2004: Supernovae as Cosmological Lighthouses, Astron. Soc. Pac., San Francisco., p.122

\bibitem[\protect\citeauthoryear{Mac Low \& Ferrara} {1999}]{mac99} Mac Low, M.-M., Ferrara, A. 1999, ApJ, 513, 142

\bibitem[\protect\citeauthoryear{Madau et al.} {1996}]{mad96} Madau, P., Ferguson, H. C., Dickinson, M. E., Giavalisco, M., Steidel, C. C., Fruchter, A. 1996, MNRAS, 283, 1388

\bibitem[\protect\citeauthoryear{Mannucci et al.} {2005}]{man05} Mannucci, F., Della Valle, M., Panagia, N., Cappellaro, E., Cresci, G., Maiolino, R., Petrosian, A., \& Turatto, M. 2005, A\&A, 433, 807

\bibitem[\protect\citeauthoryear{Marigo} {2001}]{mar01} Marigo, P. 2001, A\&A, 370, 194

\bibitem[\protect\citeauthoryear{Martin} {2005a}]{mar05a} Martin, C. L. 2005, ApJ, 621, 227

\bibitem[\protect\citeauthoryear{Martin} {2005b}]{mar05b} Martin, C. L. 2005a, in ASP Conf. Ser. 331, Extra-Planar Gas, ed. R. Brown (San Francisco: ASP), 305

\bibitem[\protect\citeauthoryear{McKee \& Ostriker} {1977}]{mck77} McKee, C. F. \& Ostriker, J. P.\ 1977, ApJ, 218, 148

\bibitem[\protect\citeauthoryear{Mo et al.} {1998}]{mo98} Mo, H. J., Mau, S., \& White, S. D. M. 1998, MNRAS, 295, 319

\bibitem[\protect\citeauthoryear{Mulchaey} {2000}]{mul00} Mulchaey, J. S. 2000, ARA\&A, 38, 289

\bibitem[\protect\citeauthoryear{Murray, Quatert, \& Thompson} {2005}]{mur05} Murray, N., Quatert, E., \& Thompson, T. A. 2005, ApJ, 618, 569 (MQT05)

\bibitem[\protect\citeauthoryear{Navarro, Frenk, \& White} {1997}]{nav97} Navarro, J. F., Frenk, C. S., \& White, S. D. M. 1997, ApJ, 490, 493

\bibitem[\protect\citeauthoryear{Oppenheimer \& Dav\'e} {2006}]{opp06} Oppenheimer, B. D. \& Dav\'e, R. A. 2006, MNRAS, 373, 1265 (OD06)

\bibitem[\protect\citeauthoryear{Papovich et al.} {2006}]{pap06} Papovich, C. et al. 2006, ApJ, 640, 92

\bibitem[\protect\citeauthoryear{P\'erez-Gonz\'alez et al.} {2005}]{per05} P\'erez-Gonz\'alez, P. G. et al. 2005, ApJ, 630, 82

\bibitem[\protect\citeauthoryear{P\'erez-Gonz\'alez et al.} {2008}]{per08} P\'erez-Gonz\'alez, P. G. et al. 2008, ApJ, 675, 234

\bibitem[\protect\citeauthoryear{Pettini et al.} {2001}]{pet01} Pettini, M., Shapley, A. E., Steidel, C. C., Cuby, J.-G., Dickinson, M., Moorwood, A. F. M., Adelberger, K. L., \& Giavalisco, M. 2001, ApJ, 554, 981

\bibitem[\protect\citeauthoryear{Pettini et al.} {2003}]{pet03} Pettini, M., Madau, P., Bolte, M., Prochaska, J.X., Ellison, S.L., \& Fan, X. 2003, ApJ, 594, 695

\bibitem[\protect\citeauthoryear{Portinari et al.} {1998}]{por98} Portinari, L., Chiosi, C., \& Bressan, A. 1998, A\&A, 334 505

\bibitem[\protect\citeauthoryear{Renzini \& Voli} {1981}]{ren81} Renzini, A. \& Voli, M. 1981, A\&A, 94, 175

\bibitem[\protect\citeauthoryear{Rozo et al.} {2007}]{roz07} Rozo, E., Wechsler, R. H., Koester, B. P., McKay, T. A., Evrard, A. E., Johnston, D., Sheldon, E. S., Annis, J., Frieman, J. A. 2007, ApJ, submitted, astro-ph/0703571

\bibitem[\protect\citeauthoryear{Rupke, Veilleux \& Sanders} {2005}]{rup05} Rupke, D. S., Veilleux, S., \& Sanders, D. B. 2005, ApJS, 160, 115

\bibitem[\protect\citeauthoryear{Ryan-Weber et al.} {2006}]{rya06} Ryan-Weber, E. V., Pettini, M., \& Madau, P. 2006, MNRAS, 371, L78

\bibitem[\protect\citeauthoryear{Salpeter} {1955}]{sal55} Salpeter E.E., 1955, ApJ, 121, 161

\bibitem[\protect\citeauthoryear{Scannapieco \& Bildsten} {2005}]{sca05} Scannapieco, E. \& Bildsten, L. 2005 ApJ, 629, L85

\bibitem[\protect\citeauthoryear{Schaerer} {2003}]{sch03} Schaerer, D. 2003, A\&A, 397, 527

\bibitem[\protect\citeauthoryear{Shapley et al.} {2003}] {sha03} Shapley, A. E., Steidel, C. C., Pettini, M., \& Adelberger, K. L. 2003, ApJ, 588, 65

\bibitem[\protect\citeauthoryear{Simcoe} {2006}]{sim06} Simcoe, R. A. 2006, ApJ, 653, 977

\bibitem[\protect\citeauthoryear{Sommer-Larson \& Fynbo} {2008}]{som08} Sommer-Larsen, J. \& Fynbo, J. P. U. 2008, MNRAS, 385, 3

\bibitem[\protect\citeauthoryear{Songaila} {2001}]{son01} Songaila, A. 2001, ApJ, 561, L153

\bibitem[\protect\citeauthoryear{Songaila} {2005}]{son05} Songaila, A. 2005, AJ, 130, 1996

\bibitem[\protect\citeauthoryear{Spergel et al.} {2007}]{spe07} Spergel, D. N. et al. 2007, ApJS, 170, 377

\bibitem[\protect\citeauthoryear{Spitoni et al.} {2008}]{spi08} Spitoni, E., Recchi, S., \& Matteucci, F. 2008, arXiv:0803.3032, accepted to A\&A

\bibitem[\protect\citeauthoryear{Springel \& Hernquist} {2002}]{spr02} Springel, V. \& Hernquist, L.  2002, MNRAS, 333, 649

\bibitem[\protect\citeauthoryear{Springel \& Hernquist} {2003a}]{spr03a} Springel, V. \& Hernquist, L.  2003, MNRAS, 339, 289 (SH03a)

\bibitem[\protect\citeauthoryear{Springel \& Hernquist} {2003b}]{spr03b} Springel, V. \& Hernquist, L.  2003, MNRAS, 339, 312 (SH03b)

\bibitem[\protect\citeauthoryear{Springel} {2005}]{spr05} Springel, V. 2005, MNRAS, 364, 1105

\bibitem[\protect\citeauthoryear{Swinbank et al.} {2007}]{swi07} Swinbank, A. M., Bower, R. G., Smith, Graham P., Wilman, R. J., Smail, I., Ellis, R. S., Morris, S. L., \& Kneib, J.-P. 2007, MNRAS, 376, 479

\bibitem[\protect\citeauthoryear{Strickland et al.} {2002}]{str02} Strickland, D. K., Heckman, T. M., Weaver, K. A., Hoopes, C. G., \& Dahlem, M. 2002, ApJ, 568, 689

\bibitem[\protect\citeauthoryear{Thielemann et al.} {1986}]{thi86} Thielemann, F.-K., Nomoto, K., \& Yokoi, K. 1986, A\&A, 158, 17


\bibitem[\protect\citeauthoryear{Twarog} {1980}]{twa80} Twarog, B. A. 1980, ApJ, 242, 242

\bibitem[\protect\citeauthoryear{Tremonti et al.} {2004}]{tre04} Tremonti, C. A. et al. 2004, ApJ, 613, 898

\bibitem[\protect\citeauthoryear{Tremonti et al.} {2007}]{tre07} Tremonti, C. A., Moustakas, J., \& Diamond-Stanic, A. M. 2007, ApJ, 663, L77

\bibitem[\protect\citeauthoryear{Trujillo et al.} {2006}]{tru06} Trujillo, I. et al., 2006, MNRAS, 373, L36

\bibitem[\protect\citeauthoryear{Trujillo et al.} {2007}]{tru07} Trujillo, I., Conselice, C. J., Bundy, K., Cooper, M. C., Eisenhardt, P., \& Ellis, R. S. 2007, MNRAS, 382, 109

\bibitem[\protect\citeauthoryear{Tsujimoto et al.} {1995}]{tsu95} Tsujimoto, T., Nomoto, K., Yoshii, Y., Hashimoto, M., Yanagida, S., \& Thielemann, F.-K. 1995, MNRAS, 277, 945

\bibitem[\protect\citeauthoryear{Wakker \& van Woerden} {1997}]{wak97} Wakker, B. P. \& van Woerden, H. 1997, ARA\&A, 35, 217

\bibitem[\protect\citeauthoryear{Wallerstein \& Knapp} {1998}]{wal98} Wallerstein, G. \& Knapp, G. R. 1998, ARA\&A, 36, 369

\bibitem[\protect\citeauthoryear{Wilman et al.} {2005}]{wil05} Wilman, R. J., Gerssen, J., Bower, R. G., Morris, S. L., Bacon, R., de Zeeuw, P. T., \& Davies, R. L. 2005, Nat, 436, 227

\bibitem[\protect\citeauthoryear{Woosley \& Weaver} {1995}]{woo95} Woosley, S. E. \& Weaver, T. A. 1995, ApJS, 101, 181

\end{thebibliography}
\end{document}